\documentclass[aps,prx,preprint,onecolumn,citeautoscript,footinbib,
eqsecnum]{revtex4-1}  
\usepackage{amsmath,amssymb,bm,bbm} 
\usepackage{graphicx}  
\usepackage{color} 
\usepackage[dvipsnames]{xcolor}
\usepackage[papersize={8.5in,11in}]{geometry}
\usepackage[colorlinks=true]{hyperref}  
\usepackage{ulem}
\usepackage[mathscr]{euscript}
\usepackage[section]{placeins}
\usepackage{comment}
\hypersetup{
    bookmarks=true,         % show bookmarks bar?
    unicode=false,          % non-Latin characters 
    pdftoolbar=true,        % show Acrobat
    pdfmenubar=true,        % show Acrobat 
    pdffitwindow=false,     % window fit to page when opened
    pdfstartview={FitH},    % fits the width of the page to the window
    pdfsubject={},   % subject of the document
    pdfcreator={},   % creator of the document
    pdfproducer={}, % producer of the document
    pdfkeywords={} {} {}, % list of keywords
    pdfnewwindow=true,      % links in new window
    colorlinks=true,       % false: boxed links; true: colored links
    linkcolor=magenta, %red,          % color of internal links (change box color with linkbordercolor)
    citecolor=blue,        % color of links to bibliography
    filecolor=magenta,      % color of file links
    urlcolor=blue           % color of external links
} 

\usepackage{amsfonts}
\usepackage{upgreek}
\usepackage{slashed}
\usepackage{latexsym}

\newcommand{\beq}{\begin{equation}}
\newcommand{\eeq}{\end{equation}}
\def\bea{\begin{eqnarray}}
\def\eea{\end{eqnarray}}
\newcommand{\nn}{\nonumber \\}

\usepackage{braket}
\usepackage{comment}
\usepackage{slashed}

\renewcommand{\vec}[1]{\boldsymbol{#1}}

\begin{document}

\preprint{\href{https://arxiv.org/abs/2001.09159}{arXiv:2001.09159}}

\title{From the pseudogap metal to the Fermi liquid using ancilla qubits}

\author{Ya-Hui Zhang}
\affiliation{Department of Physics, Harvard University, Cambridge MA 02138, USA}

\author{Subir Sachdev}
\affiliation{Department of Physics, Harvard University, Cambridge MA 02138, USA}

\date{\today
\\
\vspace{0.4in}}

\begin{abstract}
We propose a new parton theory of the hole-doped cuprates, describing the evolution from the pseudogap metal with small Fermi surfaces to the conventional Fermi liquid with a large Fermi surface. We introduce 2 ancilla qubits per square lattice site, and employ them to obtain a variational wavefunction of a fractionalized Fermi liquid for the pseudogap metal state. 
We propose a multi-layer Hamiltonion for the cuprates, with the electrons residing in the `physical' layer, and the ancilla qubits in two `hidden' layers: the hidden layers can be decoupled from the physical layer by a canonical transformation which leaves the hidden layers in a trivial gapped state. This Hamiltonian yields an emergent gauge theory which describes not only the fractionalized Fermi liquid, but also the conventional Fermi liquid, and possible exotic intermediate phases and critical points. The fractionalized Fermi liquid has hole pockets with quasiparticle weight which is large only on ``Fermi arcs'', and fermionic spinon excitations which carry charges of the emergent gauge fields.
\end{abstract}

\maketitle
\tableofcontents

\section{Introduction}
\label{sec:intro}

The structure of the pseudogap metal state in the hole-doped cuprate superconductors has long been the focus of much theoretical and experimental attention \cite{LeeWen06}. In more recent experiments, it has become clear that the pseudogap state is present only for a hole doping $p$ smaller than a critical value $p_c$ \cite{CPLT18,Vishik2012,He14,Fujita14,Badoux16,Loram01,Loram19,Michon18,Bourges18,Shen19,CPana1,CPana2,Julien19}. For $p> p_c$, many observables indicate the presence of a conventional Fermi liquid (FL) state, with a large Fermi surface enclosing a volume associated with a hole density $1+p$. While there have been many theoretical proposals for the pseudogap metal, there is as yet no framework which can capture the essential physics of {\it both} the pseudogap metal and the conventional Fermi liquid as different mean-field solutions of the same theory. Such a framework is surely needed as a starting point for a theory of the mysterious strange metal found near $p=p_c$.

We present such a framework here. We show that the introduction of 2 ancilla qubits per square lattice site leads to a valuable flexibity in describing possible correlated states of mobile electrons on the square lattice. It should be noted that the ancilla qubits are not physical degrees of freedom which can be directly observed; rather, they are theoretical tools which enable us to capture new varieties of entangled states of the electrons. 

We will describe the pseudogap metal as a fractionalized Fermi liquid (FL*) state \cite{Senthil_2003}. This state has electron-like quasiparticles around pocket Fermi surfaces enclosing a volume associated with hole density $p$. Such small pocket Fermi surfaces can appear even in the absence of any translational symmetry breaking by charge or spin density wave order. But compatibility with the Luttinger constraint requires that there be additional fractionalized spinon excitations carrying charges of an emergent gauge field \cite{Senthil_2003,SVS04,Paramekanti_2004}. Such a FL* state  is compatible with many of the experimental observations noted above for $p<p_c$, and several theoretical descriptions have been proposed \cite{WenLee96,YRZ06,YQSS10,Mei12,Punk_2012,Punk_2015,Punk17,Punk19,SS19}. However, these theories either do not describe the termination of the FL* state followed by the appearance of a FL state, or require uncontrolled non-perturbative computations to obtain a Luttinger volume violation.  
A SU(2) gauge theory \cite{SS09,DCSS15b,SCWFGS} has been proposed to described optimal doping criticality in the cuprates, but this connects naturally
to an `algebraic charge liquid' \cite{Kaul08} description of the pseudogap metal, in which the Fermi pockets are initially of spinless fermionic chargons, which have to bind with spinons to obtain the electron-like Fermi surfaces of FL* \cite{SS18}.

\begin{figure}
\includegraphics[width=4in]{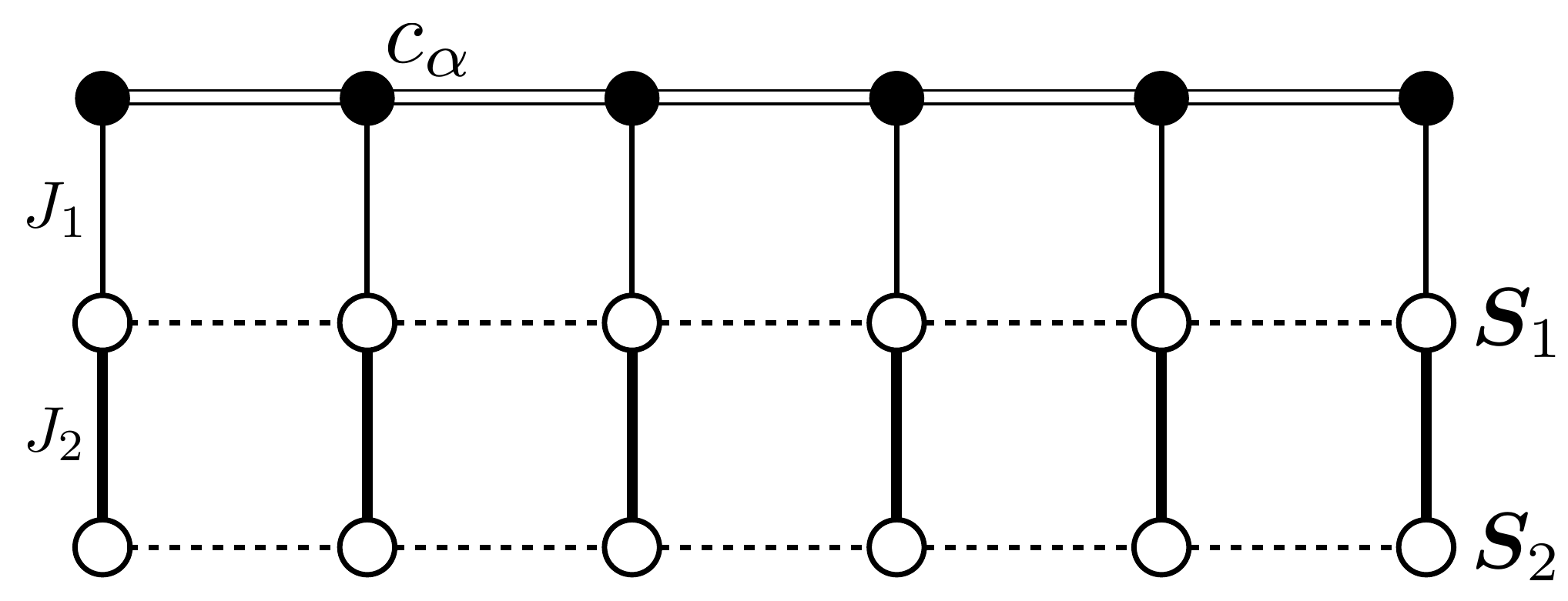}
\caption{A Hubbard model of mobile electrons $c_{i\sigma}$ of variable density $p$ in the `physical' square-lattice layer, coupled to two `hidden' square-lattice layers of ancilla qubits (spin-1/2 spins) ${\bm S}_{i;1}$ and ${\bm S}_{i;2}$. The lattice sites are labeled by $i$. We develop a theory with the exchange interactions $J_{1,2}$ finite, and take the limit $J_2 \rightarrow \infty$ of infinite antiferromagnetic exchange between the hidden layers at the end.}
\label{fig:layers}
\end{figure}

Our approach with ancilla qubits is illustrated in Fig.~\ref{fig:layers}.
We are interested in the electrons, $c_{i \sigma}$, in the physical layer, where $i$ labels a square lattice site, and $\sigma = \uparrow, \downarrow$ is the electron spin. We use the ancilla qubits in the hidden layers to generate wavefunctions and field theories for observables in the physical layer. The spin operators ${\bm S}_{i;1}$, ${\bm S}_{i;2}$ act on the qubits in the two layers. It is convenient (but not required) to represent the spins by fermions $f_{i;1\sigma}$, $f_{i;2\sigma}$ using
\beq
{\bm S}_{i;1} = \frac{1}{2} f_{i;1\sigma}^\dagger {\bm \sigma}_{\sigma\sigma'} f_{i;1 \sigma'}^{} \quad , \quad {\bm S}_{i;2} = \frac{1}{2} f_{i;2\sigma}^\dagger {\bm \sigma}_{\sigma\sigma'} f_{i;2 \sigma'}^{}
\label{eq:fermion_parton_spin}
\eeq
where ${\bm \sigma}$ are the Pauli matrices, and there must be exactly one fermion on each hidden layer site
\beq
\sum_{\sigma} f_{i;1 \sigma}^\dagger f_{i;1 \sigma} = 1 \quad, \quad \sum_{\sigma} f_{i;2 \sigma}^\dagger f_{i;2 \sigma} = 1\,.
\label{con1}
\eeq

We can now introduce our trial wavefunctions for the electrons in the physical layer. We propose a trial Slater determinant state $\left| \text{Slater}[c,f_1,f_2] \right\rangle $ for $c$, $f_1$ and $f_2$ fermions. Then we project out components of this wavefunction in which the ancilla qubits are locked in a spin singlet on each site 
\beq
\left| s \right\rangle = \prod_i \frac{1}{\sqrt{2}}(f_{i;1\uparrow}^\dagger f_{i;2 \downarrow}^\dagger - f_{i;1\downarrow}^\dagger f_{i;2 \uparrow}^\dagger) \left|0 \right\rangle\,.
\label{state_s}
\eeq
The wavefunction on the physical layer is
\beq
\left| \Phi[c] \right\rangle = \sum_a \left| a \right\rangle \left\langle a, s \right| \left. \text{Slater}[c,f_1,f_2] \right\rangle 
\label{eq:model_wave_function}
\eeq
where $\left| a \right\rangle$ is a basis of states in the physical layer. Different choices for $\left| \text{Slater}[c,f_1,f_2] \right\rangle $ will lead to different physical states $\left| \Phi[c] \right\rangle$. A FL* state is obtained by having the $c$ and $f_1$ fermions occupy linear combinations of states between the physical layer and the first hidden layer, while the $f_2$ occupy states on the second layer. The total fermion density of the $c$ and $f_1$ layers is $2-p$, and so their band structure can exhibit hole pockets of volume $p$ in the conventional Luttinger count (FL* states using an auxiliary band with total fermion density $2-p$ have also been discussed earlier \cite{YQSS10,EGMSS11}). Concomitantly the $f_2$ layer realizes the needed fractionalized spinons of the FL* state.
In the FL state, we choose  $\left| \text{Slater}[c,f_1,f_2] \right\rangle $  with decoupled the physical and hidden layers. Then the $c$ layer, with fermion density $1-p$ will have a large hole-like Fermi surface of volume $1+p$, and the hidden layers will form a trivial gapped insulator.

For analytic progress, we need the Hamiltonian of Fig.~\ref{fig:layers}, from which we can derive gauge theories. We consider the Hamiltonian
\beq
H = H_U + H_a
\eeq
where $H_U$ is a Hubbard model of electrons $c_{i \sigma}$ on the sites $i$ of a square lattice 
\beq
H_U = - \sum_{i,j} t_{ij} c_{i \sigma}^\dagger c_{j \sigma}^{} - \mu \sum_i c_{i \sigma}^\dagger c_{i \sigma}^{}
+ U \sum_{i} c_{i \uparrow}^\dagger c_{i \uparrow}^{} c_{i \downarrow}^\dagger c_{i \downarrow}^{}
\eeq 
and $H_a$ describes 2 `hidden' layers of ancilla spin $S=1/2$ qubits ${\bm S}_{i;1}$, ${\bm S}_{i;2}$ (see Fig.~\ref{fig:layers})
\beq
H_a = \frac{J_1}{2} \sum_i c_{i \sigma}^\dagger {\bm \sigma}_{\sigma\sigma'} c_{i \sigma'}^{} \cdot {\bm S}_{i;1}
+ J_2 \sum_i {\bm S}_{i;1} \cdot {\bm S}_{i;2} + H_1 ({\bm S}_{i;1}) + H_2 ({\bm S}_{i;2})
\eeq
where $H_1$ represents exchange interactions between the first hidden layer qubits ${\bm S}_{i;1}$ (which we do not specify), and similarly for $H_2$. When $J_2$ is large, the two ancilla layers will form a spin gap state; so we can safely `integrate out' the ancilla qubits, and induce near-neighbor exchange interactions between the $c_{i \sigma}$ electrons on the physical layer. Alternatively stated, a canonical transformations decouples the physical and hidden layers, at the cost of additional exchange interactions in the physical layer. Rather than accounting for these exchange interactions explicitly, more progress is possible by keeping the hidden layers `alive', and considering possible states and gauge theories in the expanded Hilbert space. In the end we can take the $J_2\rightarrow \infty$ limit (corresponding to projection onto the singlet hidden layer state $\left|s \right\rangle$ in (\ref{state_s})), which reduces the above model to the standard Hubbard model for the cuprates. A similar approach using auxiliary degrees of freedom has been taken by Refs.~\onlinecite{PH98,read1998lowest} for a description of the bosonic composite Fermi liquid in the lowest Landau level.  

The outline of the rest of the paper is follows. We present a gauge theory description of phases of $H$ in Section~\ref{sec:gauge}. This leads to a mean-field description of the FL* and FL phases. The physical properties of the FL* phase, including its photoemission spectrum, are described in Section~\ref{sec:pseudogap}. The critical region between the FL* and FL phases is discussed in Section~\ref{sec:critical}, including a possible intermediate phase, or a direct transition. Some directions for future research are noted in Section~\ref{sec:future}, and we summarize in Section~\ref{sec:summary}.

\section{Gauge structure and Mean field theory}
\label{sec:gauge}

\subsection{$(SU(2)_1\times SU(2)_2 \times SU(2)_S)/Z_2$ gauge structure}

The spins in the hidden layers can be represented by the standard fermionic partons \cite{LeeWen06} in (\ref{eq:fermion_parton_spin}). Naively we can just form mean field theories using $c,f_1,f_2$. However, this kind of analysis does not incorporate the large $J_2$. For example, let us consider mean field ansatz for which $c$ decouples from $f_1,f_2$; then $f_1,f_2$ can form gapless spin liquids. However, in the large $J_2$ region, the spin carried by $f_1$ and $f_2$ must be gapped. This is similar to a ``Mott'' gap in spin channel. In the familiar Hubbard model, we use slave boson theory to describe the Mott transition and incorporate the charge gap at large $U$.  Here we can use a similar slave spin approach to incorporate the spin gap at large $J_2$. 

Therefore , we perform a further fractionalization of $f_1$ and $f_1$ in (\ref{eq:fermion_parton_spin}):
\begin{equation}
    f_{i; a\sigma}=R_{i; \sigma \tilde \sigma}\tilde f_{i;a\tilde \sigma}
\end{equation}
where $a=1,2$ and the slave spin $R$ is a $SU(2)$ matrix, similar to that introduced in Ref.~\onlinecite{SS09}. Basically this means the ``spin'' index $\tilde \sigma$ carried by $\tilde f$ can be freely rotated by a $SU(2)$ gauge transformation:
\begin{equation}
   \left( \begin{array}{c} 
    \tilde f_{i;a\uparrow} \\
    \tilde f_{i;a\downarrow}
    \end{array}\right ) \rightarrow
     U_{i;S} \left( \begin{array}{c} 
    \tilde f_{i;a\uparrow} \\
    \tilde f_{i;a\downarrow}
    \end{array}\right )
    \label{eq:gaugeS}
\end{equation}
where $U_{i;S} \in SU(2)$.  Accordingly the slave spin  transforms as $R_i \rightarrow R_i U^\dagger_{i;S}$.  We label this gauge transformation as $SU(2)_S$. See also Appendix~\ref{app:constraints} for further discussion on the origin of this expanded gauge structure.
% Higgsing or confinement of the $SU(2)_S$ gauge field will lead to the needed spin gap.

It is well known \cite{LeeWen06} that the  parton representation for each layer  in (\ref{eq:fermion_parton_spin}) has another  $SU(2)$ gauge transformation in the particle-hole channel:
\begin{equation}
    \left(
    \begin{array}{c}
    \tilde f_{i;a\uparrow}\\
    \tilde f_{i;a\downarrow}^\dagger
    \end{array}\right)\rightarrow U_{i;a}\left(
    \begin{array}{c}
    \tilde f_{i;a\uparrow}\\
    \tilde f_{i;a\downarrow}^\dagger
    \end{array}\right)
    \label{eq:gauge12}
\end{equation}
where $a=1,2$ and $U_{i;a}\in SU(2)$.  The $U_{i;1}$ and $U_{i;2}$ are two independent gauge transformations for $\tilde f_1$ and $\tilde f_2$ respectively, which commute with $U_{i,S}$ \cite{XS10}. The slave spin $R_{i}$ remains unchanged under both $U_{i;1}$ and $U_{i;2}$.   We label these two gauge transformations as $SU(2)_1$ and $SU(2)_2$ respectively.  In total the gauge structure of our parton theory in terms of $\tilde f_{i;a\sigma}$ is $(SU(2)_1\times SU(2)_2 \times SU(2)_S)/Z_2$.  Here we need to mod out the transformation: $\tilde f_{i;a\sigma}\rightarrow - \tilde f_{i;a\sigma}$. The combined transformations in 
(\ref{eq:gaugeS}) and (\ref{eq:gauge12}) are similar to the O(4) fractionalization of Ref.~\onlinecite{XS10}, but the $SU(2)_{S}$ transformations for $a=1,2$ fermions have been tied to each other by the large $J_2$.

The fermions $\tilde f_{1}, \tilde f_{2}$ are neutral under both physical charge and spin probes because they couple to the $(SU(2)_1 \times SU(2)_2 \times SU(2)_S)/Z_2$ gauge fields.  If the boson $R$ is gapped, and there is no further Higgs term, in the strong gauge field coupling limit $\tilde f_1, \tilde f_2$ will be confined to form on-site spin singlets.  A more interesting possibility can be obtained by a mean field ansatz in which we Higgs the gauge field by coupling $\tilde f_{a;\sigma}$ to the physical electron $c$: we will explore such an ansatz in the remainder of the paper.

\subsection{Mean field theory}

 Let us define $C=(c_\uparrow, c_\downarrow, c^\dagger_\downarrow, -c^\dagger_\uparrow)^T$, $\Psi_1=(\tilde f_{1\uparrow}, \tilde f_{1\downarrow}, \tilde f^\dagger_{1\downarrow}, -\tilde f^\dagger_{1\uparrow})^T$ and  $\Psi_2=(\tilde f_{2\uparrow}, \tilde f_{2\downarrow}, \tilde f^\dagger_{2\downarrow}, -\tilde f^\dagger_{2\uparrow})^T$.  We define $\rho_a$ and $\mu_a$ as Pauli matrices acting on the spin and particle-hole channels respectively.  Then the $SU(2)_S$ gauge transformation $U_{i;S}$ is generated by $\rho_a$, and acts on both $\Psi_1$ and $\Psi_2$. The $SU(2)_1$ gauge transformation $U_{i;1}$ is generated by $\mu_a$, and acts only on $\Psi_1$. Similarly, the $SU(2)_2$ gauge transformation $U_{i;2}$ is generated by $\mu_a$ and acts only on $\Psi_2$. In this basis, the gauge transformations $U_{i;1}$, $U_{i;2}$ and $U_{i;S}$ are $4\times 4$ matrices.

After condensation of approriate Higgs fields, we obtain our proposed mean field theory:
\beq
H_M = \sum_i \left(  C_{i}^\dagger \bold B_{i;1}  \Psi_{i; 1} + \mbox{H.c.} + \Psi_{i;1}^\dagger \bold B_{i;2} \Psi_{i;2} + \mbox{H.c.} \right) +H_C+ H_1+H_2+H_R
\label{mean_field_theory}
\eeq
where $\bold B_{i;1}$ and $\bold B_{i;2}$ are $4\times 4$ matrices. The Hamiltonian $H_C$ is the bare kinetic term for physical electron $c$. The Hamiltonians $H_1$ and $H_2$ are mean field ansatzes for $\Psi_1$ and $\Psi_2$ which we will specify later. The Hamiltonian $H_R$ is the ansatz for the slave spin $R_i$; at large $J_2$ limit, we can assume $R$ is massive and $\langle R_i \rangle=0$ by invariance under physical global spin rotations.

Under the $(SU(2)_1\times SU(2)_2 \times SU(2)_S)/Z_2$ gauge transformation, $\bold B_{i;1}$ and $\bold B_{i;2}$ transform as
\begin{align}
\bold B_{i;1} & \rightarrow  \bold B_{i;1} U_{i;1}^\dagger U_{i;S}^\dagger \notag\\
\bold B_{i;2} & \rightarrow U_{i;S} U_{i;1} \bold B_{i;2}U^\dagger_{i;2} U_{i;S}^\dagger
\end{align}
Note that the $U_{i;a}$ commute with the $U_{i;S}$ \cite{XS10}, and so their ordering is unimportant.

We also have a global $U(2)_C$ symmetry corresponding to charge conservation and spin rotation. In the basis of $C$, the $U(1)$ part is generated by $\mu_z$ while the $SU(2)$ spin rotation is generated by $\vec \rho$.  Under this global $U(2)_C$ transformation $U_C$, $C_i \rightarrow U_C C_i$, $\bold B_{i;1} \rightarrow U_C \bold B_{i;1}$, while $\bold B_{2}, \Psi_1, \Psi_2$ remain unchanged.  Hence $\bold B_{i;1}$ carries both physical charge-spin and gauge charges.

The mean-field characterizations of the phases are 
\bea
{\rm FL^*}&:&~ \langle \bold B_1 \rangle =\Phi \rho_0 \otimes \mu_z, \quad \langle \bold B_2 \rangle = 0 \nn
{\rm FL}&:&~ \langle \bold B_1 \rangle = 0, \quad \langle \bold B_2 \rangle =\Phi' \rho_0 \otimes \mu_z
\label{e6}
\eea
where $\Phi$  and $\Phi'$ are  real numbers. Here we chose a specific gauge. Equivalent ansatzes can be obtained through gauge transformations.  In the FL phase, $\langle \bold B_2 \rangle \neq 0$ does not need to be put by hand; once $\langle \bold B_1 \rangle =0$ and $\langle R_i \rangle=0$ , gauge fluctuation can confine $\Psi_1$ and $\Psi_2$ automatically, which is equivalent to the effect of non-zero $\langle \bold B_2 \rangle$.

The condensate $\langle \bold B_1 \rangle =\Phi \rho_0 \otimes \mu_z $ higgses the $(SU(2)_1 \times SU(2)_2 \times SU(2)_S)/Z_2$ gauge fields (the hopping terms in $H_1$ and $H_2$ also Higgs parts of the gauge symmetry even with $\langle \bold B_1 \rangle=0$). In a more precise language, it locks  the $SU(2)_C \subset U(2)_C$ background field to  the internal gauge fields corresponding to $SU(2)_S$.  Thus the spin index  $\tilde \sigma$ carried by $\tilde f_{1\tilde \sigma}$ and $\tilde f_{2\tilde \sigma}$ can now be identified as a physical spin index.  The internal $U(1)$ gauge field generated by $\mu_z$ in $SU(2)_1$ is locked to the physical electromagnetic field; after the condensation of $\bold B_1$, $\tilde f_1$ can be viewed as electron and $\tilde f_2$ can be identified as a spinon. The fermion $\tilde f_{2}$ is still charge neutral because $SU(2)_2$ is not locked to the physical background field.

\section{Properties of the pseudogap metal}
\label{sec:pseudogap}

In this section we provide details of the FL* mean field ansatz for the pseudogap metal in the underdoped region and discuss its properties. 

When $\langle \bold B_1 \rangle =\Phi \rho_0 \otimes \mu_z$, $\langle \bold B_2 \rangle=0$, we have the following mean field theory
\begin{equation}
    H_M=H_{c,\tilde f_1}+H_{\tilde f_2}
\end{equation}
The Hamiltonian $H_{c,\tilde f_1}$ describes the electron Fermi surface while $H_{\tilde f_2}$ describes the phase of the spinon $\tilde f_2$.  Small Fermi surfaces with Luttinger volume $A_{FS}=p/2$ can be obtained by 
\begin{align}
    H_{c,\tilde f_1}&=\sum_{ij}(-t_{c;ij}c^\dagger_{i\sigma} c_{j\sigma} + t_{1;ij} \tilde f^\dagger_{i;1\sigma} \tilde f_{j;1 \sigma}+h.c.) -\mu_c \sum_i c^\dagger_{i;\sigma} c_{i;\sigma} -\mu_1 \sum_i \ \tilde f^\dagger_{i;1\sigma}\tilde f_{i;1\sigma} \nonumber \\
    &~~~~~~~~~~ +\Phi\sum_i (c^\dagger_{i;\sigma} \tilde f_{i;1\sigma}+h.c.)
    \label{eq:mean_field_c_f1}
\end{align}
where $\mu_1$ is added to fix $n_{\tilde f_1}=\sum_\sigma \langle \tilde f^\dagger_{i;1\sigma} \tilde f_{i;1\sigma} \rangle=1$, and $\mu_c$ fixes $n_c=1-p$. For the hopping parameters $t_{c;ij}$, we use $t=1, t'=-0.22, t''=0.19$ which can reproduce the shape of the Fermi surface in the overdoped regime.   The hopping parameters for $\tilde f_{1}$ should be determined by minimizing the energy for the wavefunction in (\ref{eq:model_wave_function}). Alternatively, we can also view them as phenomenological parameters and fit them from experimental data.   Here we choose $t_1=1, t'_1=-0.1, t''_1=0.1$ for the purpose of illustration. In practice, the hopping parameters $t_{1;ij}$ can also have dependences on doping level $p$. Later we will provide an intuitive explanation why the hoppings of $\tilde f_{1}$ have the opposite sign to that of $c$.

For the spinon $\tilde f_2$, we use the familiar $d$ wave pairing ansatz \cite{LeeWen06}:
\begin{equation}
    H_{\tilde f_2}=-t_2 \sum_{\langle ij \rangle} \tilde f^\dagger_{i;2}\tilde f_{j;2}+  \sum_{ i, \hat{\mu}=x,y } \Delta_{\hat{\mu}}(\epsilon_{\sigma \sigma'}\tilde f^\dagger_{i;2 \tilde \sigma}\tilde f^\dagger_{i+\hat{\mu};2\tilde \sigma'}+h.c.)
    \label{eq:mean_field_f2}
\end{equation}
where, $\Delta_{\hat x}=-\Delta_{\hat y}=\Delta$.  This ansatz is equivalent to the staggered flux ansatz, and the spinon $\tilde f_2$ is in a $U(1)$ Dirac spin liquid phase.

\begin{figure}[ht]
\centering
\includegraphics[width=0.95 \textwidth]{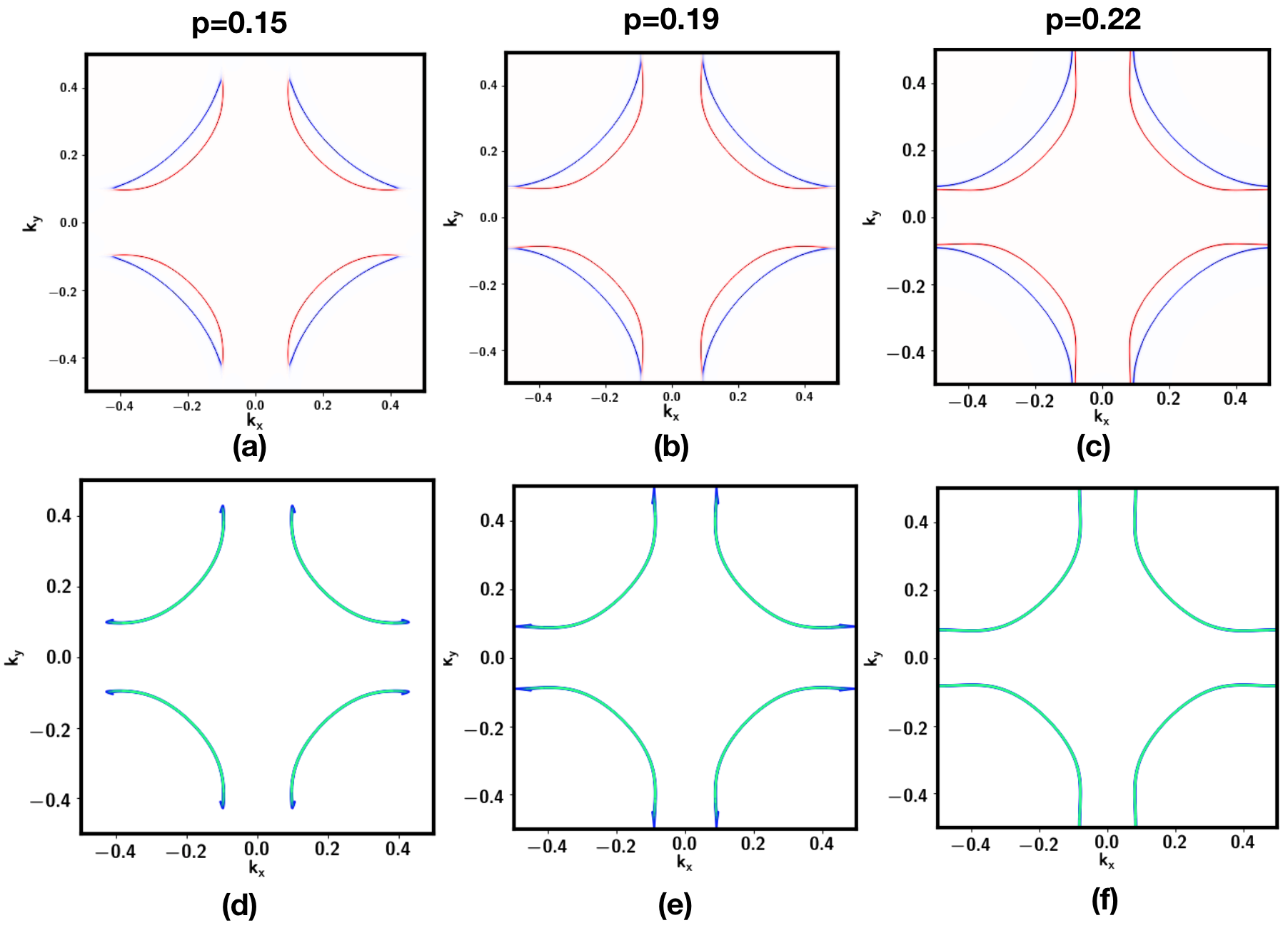}
\caption{Spectral function $A(\omega=0, \mathbf k)$ from mean field theory at various doping levels. $k_x$ and $k_y$ is in unit of $\frac{2\pi}{a}$. In (a-c),  red line is dominated by $A_c(\omega=0, \mathbf k)$ while blue line is dominated by $A_{\tilde f_1}(\omega=0,\mathbf k)$. In (d-f) we only show $A_c(\omega=0, \mathbf k)$, which clearly shows "Fermi arc" at $p<p_l\approx 0.19$. At $p_l \approx 0.19$, there is a Lifshitz transition. }
\label{fig:fermi_arc}
\end{figure}

The gap at the anti-node is opened by $\Phi \neq 0$.  We choose to use $\Phi(p)=0.25 \sqrt{0.23-p}$.  Then we can calculate spectral densities $A_c(\omega, \mathbf k)=({1}/{\pi}) \text{Im}  \langle c^\dagger(\omega, \mathbf k) c(\omega, \mathbf k) \rangle$ and $A_{\tilde f_1}(\omega, \mathbf k)=({1}/{\pi}) \text{Im}  \langle \tilde f^\dagger_1(\omega, \mathbf k)\tilde f_1(\omega, \mathbf k) \rangle$. We show plots of the calculated spectral densities in Fig.~\ref{fig:fermi_arc}.  Between $p_l\approx 0.19$ and $p_c=0.23$, the anti-node is not gapped even if $\Phi \neq 0$. Instead, there is one hole Fermi surface dominated by $c$ and one electron pocket dominated by $\tilde f_1$.  The total Hall number is close to $p$ and this region still belongs to the pseudogap phase. When we decrease $p$ below $p_l$, there is a Lifshitz transition and the Fermi surfaces are reconstructed to four small hole pockets centering at node $\mathbf{K_N}=(\pm \frac{\pi}{2},\pm \frac{\pi}{2})$.  For each pocket, one side is dominated by $\tilde f_1$ and thus has almost vanishing spectral weight in terms of $c$. As a result, only Fermi arcs are visible in an ARPES measurement. 
Note also the similarity of Fig.~\ref{fig:fermi_arc} to the STM observations in Refs.~\onlinecite{He14,Fujita14}.

\begin{figure}[ht]
\centering
\includegraphics[width=0.95 \textwidth]{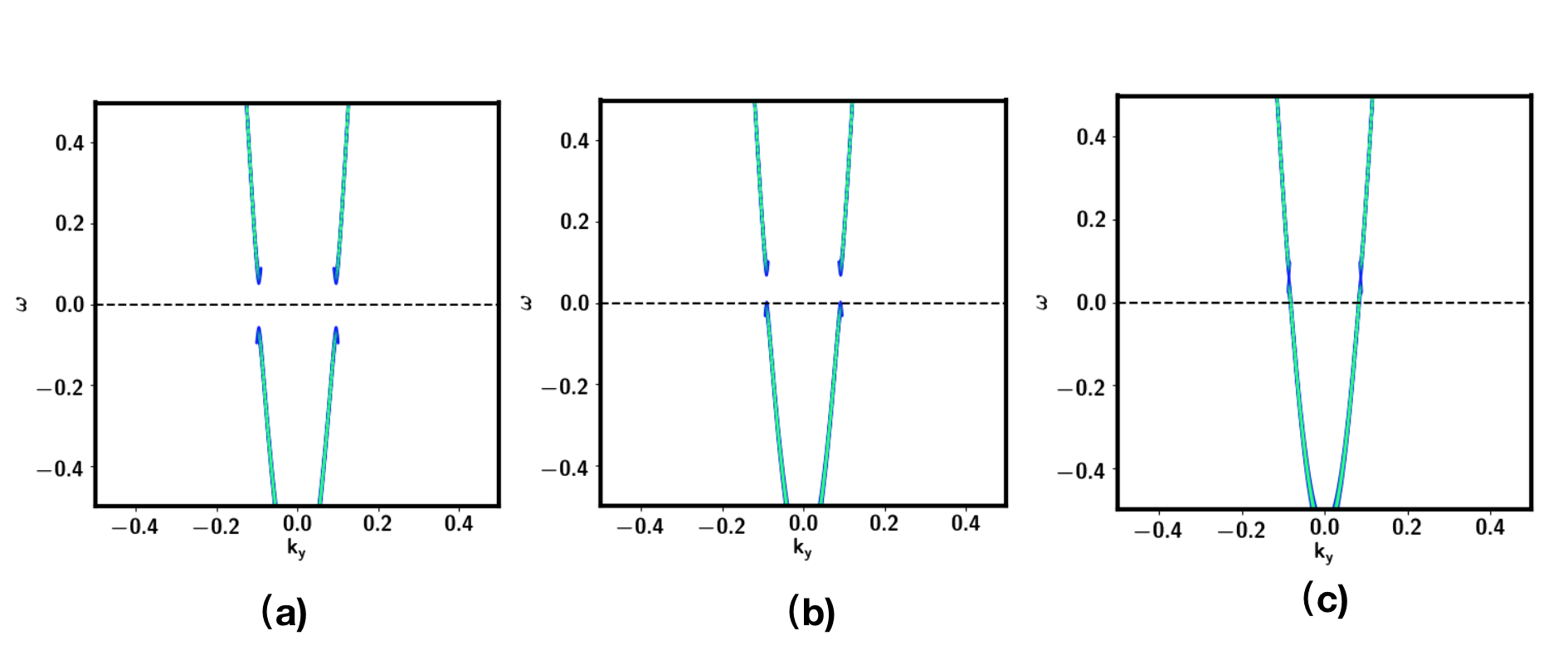}
\caption{Spectral function $A_c(\omega, \mathbf k=(\pi, k_y))$ calculated from mean field theory at various doping levels. The doping is (a) $p=0.15$; (b) $p=0.19$ and (c) $p=0.21$. In ARPES only the $\omega<0$ region can be measured at low temperature. Because of the gauge fluctuations at high energy, the true spectral function may be quite different and no sharp quasi-particle peak exists. }
\label{fig:anti_node_spectrum}
\end{figure}

In Fig.~\ref{fig:anti_node_spectrum} we show $A_c(\omega, \mathbf k=(\pi,k_y))$. We define anti-node to be at $\mathbf K_{AN}=(\pi,\pm \delta)$ which separates the $n_c(\mathbf k)=1$ and $n_c(\mathbf k)=0$ regions along the cut of $k_x=\pi$.  Then at $\mathbf k=\mathbf K_{AN}$, $A_c(\omega, \mathbf K_{AN})$ has a peak at $\omega=-\Delta$.    Here the sharp quasi particle peak at $\omega=-\Delta$ is an artifact of the mean field calculation which ignores gauge fluctuation. Even inside the pseudogap phase, the mass of the higgsed gauge fields is at order of $\Phi\sim \Delta$.  For high energy region $|\omega|>\Delta$, gauge fluctuations can not be ignored and may completely destroy the quasi-particle peak at anti-node.   In contrast, around the node, the Fermi arc is at zero energy and the gauge fluctuations do not have strong influences on spectral function here. 

\begin{figure}[ht]
\centering
\includegraphics[width=0.95 \textwidth]{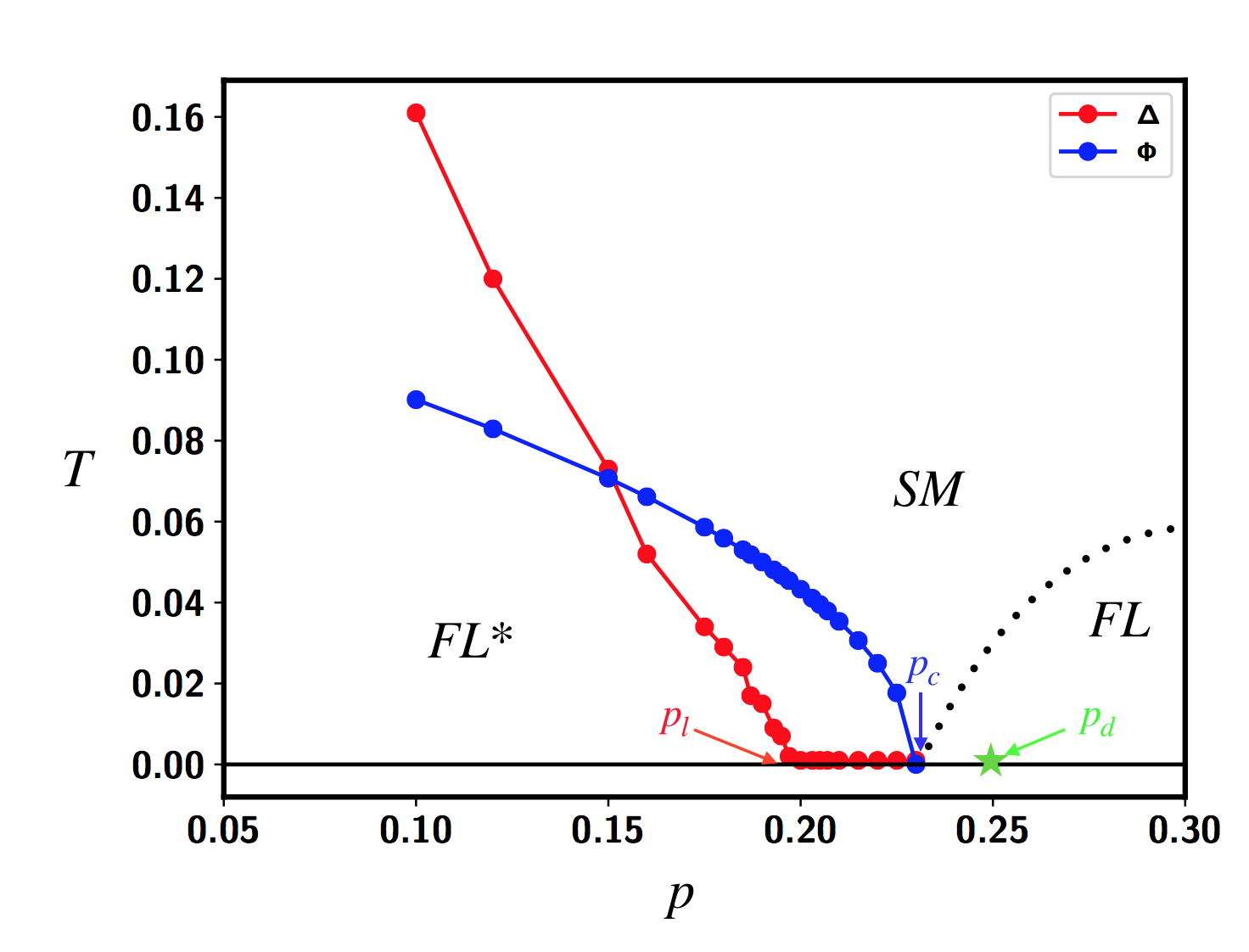}
\caption{Phase diagram in $T-x$ space. We choose $\Phi=0.25 \sqrt{0.23-p}$. $\Delta$ is calculated from mean field theory and is in unit of $t=1$. The dashed line is drawn by hand to show the other crossover line of the quantum critical region. }
\label{fig:phase_diagram}
\end{figure}

We define $\Delta$ as the gap of the anti-node.  The dependence of $\Delta$ and $\Phi$ on the doping level is shown in Fig.~\ref{fig:phase_diagram}. The doping $p_c$ is the true quantum critical point $QCP$ at the end of the pseudogap phase. For our parameters, the anti-node gap $\Delta$ onsets at a smaller doping $p_l$ through a Lifshitz transition. In slightly overdoped region $p_c<p<p_d$, there may be ghost Fermi surfaces decoupled from the Fermi liquid, which get confined and disappear at a larger doping $p_d$.

\subsection{Physical meaning of auxiliary fermions}

At the $J_2 \rightarrow \infty$ limit, the wavefunction of (\ref{eq:model_wave_function}) can be written as
\begin{equation}
    \ket{\Phi}=\left(\braket{s| \text{Slater} [c,\tilde f_1]  \text{Slater} [\tilde f_2] }\right)\ket{s}
\end{equation}
where $\ket{s}= \prod_i (\tilde f_{i;1\uparrow}^\dagger \tilde f_{i;2 \downarrow}^\dagger - \tilde f_{i;1\downarrow}^\dagger \tilde f_{i;2 \uparrow}^\dagger)/\sqrt{2} \left|0 \right\rangle$. $\text{Slater} [c,\tilde f_1]$ is the ground state of $H_{c,\tilde f_1}$ in Eq.~\ref{eq:mean_field_c_f1}. $\text{Slater} [\tilde f_2]$ is the ground state of $H_{\tilde f_2}$ in (\ref{eq:mean_field_f2}). 

The above wavefunction is a state purely in the Hilbert space of the physical layer.  Thus $\tilde f_1$ and $\tilde f_2$ should correspond to physical degrees of freedom. A natural question is: what is the physical meaning of these auxiliary fermions inside the physical Hilbert space?  To gain intuition, let us look at the zero doping case first. In this case, $\Phi$ can gap out both $c$ and $\tilde f_1$ and we have a Mott insulator. Therefore we should interpret $\Phi$ as Mott gap. The $\Phi$ condensate binds electron $c^\dagger$ to hole $\tilde f_1$, and so we should view $\tilde f_1$ as creation operator for a ``correlation hole''.  This correlation hole may be quite nonlocal and should not be confused with on-site hole operator $c_i$.  In a certain sense, the physics is similar to that of fractional quantum Hall effect (FQHE). In the FQHE system, because of Coulomb interaction, electron also binds with its correlation hole (also called a vortex) and only the the bound state (a composite fermion) can move coherently.  In our case, the Hubbard $U$ also favors the binding between electron and correlation hole, which causes Mott localization.  Unlike the FQHE, here in (\ref{eq:mean_field_c_f1}) we use a BCS description of the exciton binding instead of viewing the exciton as the fundamental particle.  The exciton binding means that the exciton $c^\dagger \tilde f_1$ moves coherently, and thus the hopping of $\tilde f_1$ should be similar to that of $c^\dagger$. Thus it is natural that the hopping of $\tilde f_1$ has the opposite sign to that of $c$.

At zero doping, $c,\tilde f_1$ are gapped and generate the upper Hubbard band, and the lower Hubbard band for the Mott insulator.  The fermion $\tilde f_2$ can be identified as spinon at low energy.  At small doping $p$, the ``Mott gap'' does not close immediately and the doped hole just enters the lower Hubbard band, and forms small hole pockets. In this sense, the pseudogap is inherited from the Mott gap of the undoped parent compound.

\section{Critical theory}
\label{sec:critical}

In this section we provide the theory at the critical point $p_c$, where a non-zero $\Phi$ onsets. At the critical point, the Higgs condensate $\mathbf B_1$  fluctuates along the manifold generated by the gauge transformations. The hopping terms for $\tilde f_1$ and $\tilde f_2$ in (\ref{eq:mean_field_c_f1}) and (\ref{eq:mean_field_f2}) break the $(SU(2)_1\times SU(2)_2 \times SU(2)_S)/Z_2$ down to $(U(1)_1 \times U(1)_2 \times SU(2)_S)/Z_2$. Basically there is no gauge transformation rotating a particle to a hole. As a consequence, the fluctuation of $\bold B_1$ is generated by a $U(2)=(U(1)_1 \times SU(2)_S)/Z_2$ transformation.  We define $\tilde C_i=(c_{i;\uparrow},c_{i;\downarrow})^T$, $\tilde \Psi_{i;1}=(\tilde f_{i;1\uparrow},\tilde f_{i;1\downarrow})^T $ and $\tilde \Psi_{i;2}=(\tilde f_{i;2\uparrow}, \tilde f_{i;2\downarrow})^T$. We define the $U(1)$ gauge fields for $U(1)_1$ and $U(1)_2$ as $a_1$ and $a_2$.  The $SU(2)_S$ gauge field is labeled as $\vec{\alpha}$.

The critical theory is 
\begin{align}
L=L_C+L_{\Psi_1, a_1, \vec \alpha}+L_{\Psi_2,a_2, \vec \alpha}+L_{B,a_1, \vec \alpha}+ (\tilde C^\dagger(\tau,\mathbf x) B (\tau, \mathbf x) \tilde \Psi_1(\tau, \mathbf x)+h.c.)
\end{align}
where $B_1(\tau,\mathbf x)$ is a $2 \times 2$ matrix parameterized as $B_1(\tau,\mathbf x)=\Phi(\tau,\mathbf x) U(\tau,\mathbf x)$. Here $\Phi(\tau,\mathbf x)$ is a complex field and $U(\tau,\mathbf x)$ is a $SU(2)$ matrix field.

The Lagrangian $L_C$ is the action for the fermi liquid theory of physical electron. $L_{\Psi_2, a^2, \vec \alpha}$ is the action for the spinon coupled to $(U(1)_2 \times SU(2)_S)/Z_2 $ gauge field.  For our Dirac spin liquid ansatz, it has $N=2$ Dirac fermions in the fundamental representation coupled to the $U(2)$ gauge field.

The theory for the critical boson $B$ is
\begin{equation}
    L_{B,a^1,\vec \alpha}= \frac{1}{g_B} |\partial_\mu B_{\alpha \beta} -a^1_\mu B_{\alpha \beta} - \alpha^a_\mu B_{\alpha \beta'} \rho^a_{\beta' \beta}|^2-m |B_{\alpha \beta}|^2
\end{equation}
which is a non-linear sigma model coupled to $U(2)$ gauge field. We used Einstein summation for $\alpha, \beta, \beta'=\uparrow, \downarrow$.

The fermion $\Psi_1$ forms a Fermi surface and couples to both $a^1$ and $\vec \alpha$:
\begin{equation}
    L_{\Psi_1, a^1, \vec \alpha}=  \Psi^\dagger_1 (\partial_\tau -a^1_0 -\alpha^a_0 \rho^a)\Psi_1-\frac{\hbar^2}{2 m^*} \Psi^\dagger_1(\partial_i-a^1_i-\alpha^1_i \rho^a)^2\Psi_1
\end{equation}

The quantum critical point is tuned by the mass term $m$ for the critical boson. At $m<0$, $\langle B  \rangle = \Phi \rho_0$ and this is a FL* phase.  When $m>0$, $\tilde \Psi_1$ and $\tilde \Psi_2$ decouple from the physical electron.  In this case $\tilde \Psi_1$ forms ghost a Fermi surface which couples to neither charge nor spin probes.  $\tilde \Psi_1$ couples to a $U(1)$ and a $SU(2)$ gauge field. It is known that $U(1)$ gauge field suppresses pairing, while $SU(2)$ gauge field induces pairing \cite{MMSS15}.  Thus it is not clear whether this ghost Fermi surface is stable or not.  If the ghost Fermi surface is unstable with infinitesimal $m>0$, there is hope to have a direct transition between  the FL* phase and the FL phase. Otherwise the ghost Fermi surface can survive until a larger doping $p_d>p_c$.  In any case, our theory implies that there is a ghost Fermi surface  coexisting with the physical Fermi surface in the strange metal region and the density of states (DoS) should be significantly larger than that of the overdoped Fermi liquid. A recent experimental measurement of specific heat indeed found that $\gamma=C/T$ close to critical region is almost four times larger  that that of the 
overdoped FL \cite{Michon18}.  An independent measurement of the effective mass is needed to subtract the contribution from the Fermi liquid part and test the existence of ghost Fermi surface.

\section{Future directions}
\label{sec:future}

There are several directions to generalize our theory. In this paper we restricted our analysis to the symmetric FL* phase, but it is easy to incorporate symmetry breaking orders. For example, we can let the ansatz of $\tilde f_1$ have nematic order or loop current order. In this case, the nematic order or loop current order will onset when $\Phi \neq 0$, and coincides with the onset of the pseudogap phase. However, the symmetry breaking order is just a byproduct of the pseudogap phase and does not play any essential role.   Alternatively, $\tilde f_2$ can be put in an ansatz with antiferromagnetic order. Thus the theory describes evolution from an antiferromagnetic metal with small Fermi surfaces towards the Fermi liquid phase with large Fermi surface.  In contrast to the standard Hertz-Millis theory of antiferromagnetic critical point,  this theory allows a jump of carrier density across the critical point  and  may be relevant for the quantum critical point in heavy fermion systems.

Our framework can also be easily generalized to $SU(N)$ Hubbard model with any $N$.  At integer filling $n_c$, there is a Mott insulator with $n_c$ electrons per site. We also introduce an auxiliary fermion $\tilde f_1$ at density $n_{f_1}=N-n_c$ and an auxiliary fermion $\tilde f_2$ with density $n_{\tilde f_2}=n_c$. The fermions $\tilde f_1$ and $\tilde f_2$ can again form trivial $SU(N)$ singlet per site, and there is a $U(1)_1 \times U(1)_2 \times SU(N)_S$ gauge structure.  Then upon doping at filling $n=n_c-x$, we can have a coupling like $-\Phi c^\dagger_\alpha \tilde f_{1\alpha}$, which can lead to small Hall number $\eta_H=-x$. The fermion $\tilde f_2$ can be viewed as a spinon, and be put in spin liquid phase or ordered phase.  This implies that pseudogap metal is a quite universal phenomena upon doping a generic Mott insulator.  Recently an approximate $SU(4)$ Hubbard model is shown to be realized in graphene moir\'e superlattice \cite{chen2019evidence,zhang2019bridging} and thus this generalized theory may be relevant there.

\section{Summary}
\label{sec:summary}

We have proposed a new framework for describing the pseudogap phase obtained from doping a Mott insulator, which can also be extended towards understanding its evolution towards the conventional Fermi liquid at larger doping. 
We showed that the use of ancilla qubits allows us to address the complete doping evolution in a single mean-field framework, which has not been possible in previous work.
We applied our theory to the hole doped cuprates. At small doping, we provide a simple parton mean field theory of the fractionalized Fermi liquid (FL*) with small Fermi surfaces and reproduce the ``Fermi arc'' in ARPES experiments. We also provide a critical theory at the end of the pseudogap phase across which the carrier density jumps from $p$ to $1+p$. Our theory finds a ``ghost'' Fermi surface close to the optimal doping which should significantly enhance the density of states.

Finally, we note the relationship of the present gauge theory of optimal doping criticality to a recent $SU(2)$ gauge theory \cite{SS19,SPSS19} of the same regime. The common features are a $SU(2)_S$
gauge field and a large Fermi surface of gauge-neutral electrons $c_\sigma$. The differences are that the other theory \cite{SS19,SPSS19} has ({\it i\/}) multiple Higgs fields that transform in the {\it adjoint\/} of $SU(2)_S$, ({\it ii\/}) the Higgs fields transform non-trivially under the space group of the square lattice, ({\it iii\/}) bosonic spinon excitations which remain gapped across the transition. In contrast, our present theory has ({\it i\/}) a single Higgs field that transforms as a $SU(2)_S$ {\it fundamental\/}, and also under a separate emergent $U(1)$ gauge field, ({\it ii\/}) the Higgs fields is also a fundamental of the global spin rotation $SU(2)$ and the electromagnetic $U(1)$, ({\it iii\/}) there are gapless ``ghost'' fermionic excitations which carry neither spin nor charge, but which carry the charges of the $SU(2)_S$ gauge field, the emergent $U(1)$ gauge field, and another $SU(2)_2$ gauge field. For our present theory, it is possible to take a linear combination of the emergent and electromagnetic $U(1)$'s and transfer the electromagnetic charge from the Higgs field to a ghost fermion \cite{IoffeLarkin89}. 

\appendix

\section{Constraints and gauge structure}
\label{app:constraints}

We make some additional comments here on the origin of the $(SU(2)_1\times SU(2)_2 \times SU(2)_S)/Z_2$ gauge structure.

The complete constraints from the single occupancy of the hidden layers are (generalizing (\ref{con1})) \cite{LeeWen06}:
\bea
\sum_{\tilde \sigma} \tilde f_{i;1 \tilde \sigma}^\dagger \tilde  f_{i;1 \tilde \sigma} = 1 \quad &,& \quad \sum_{\tilde \sigma} \tilde f_{i;2 \tilde \sigma}^\dagger \tilde  f_{i;2 \tilde \sigma} = 1, \nonumber \\
\sum_{\tilde \sigma \tilde \sigma'} \epsilon_{\tilde \sigma \tilde \sigma'} \tilde  f_{i;1 \tilde\sigma} \tilde f_{i;1 \tilde \sigma'} = 0 \quad &,& \quad \sum_{\tilde \sigma \tilde \sigma'} \epsilon_{\tilde \sigma \tilde \sigma'} \tilde f_{i;2 \tilde\sigma} \tilde f_{i;2  \tilde \sigma'} = 0\,.
\label{con2}
\eea
We also have the Hermitian conjugates of the constraints in the second line, and so (\ref{con2}) contains a total of 6 real constraints on each lattice site $i$. In the limit $J_2 \rightarrow \infty$, the spins in the hidden layers are projected onto singlets on each site, and so we have 3 additional constraints
\beq
{\bm S}_{i;1} + {\bm S}_{i;2} = 0\,,
\label{con3}
\eeq
after replacing the $f_\sigma$ in (\ref{eq:fermion_parton_spin}) by $\tilde f_{\tilde\sigma}$.
We note here that these 9 constraints per site correspond precisely to the 9 generators of the $(SU(2)_1\times SU(2)_2 \times SU(2)_S)/Z_2$ gauge symmetry.

As discussed in Ref.~\onlinecite{LeeWen06} (see also Ref.~\onlinecite{MSSS18}), the constraints in (\ref{con2}) can be expressed as the vanishing of the Nambu pseudospin operator for each $i$ and $a=1,2$. This pseudospin operator transforms as an adjoint under 
$SU(2)_1 \times SU(2)_2$, and so its vanishing is maintained under these transformations. The expressions in (\ref{con2}) are explicitly spin rotation invariant, and so are also invariant under the $SU(2)_S$ transformation in (\ref{eq:gaugeS}).

Another feature of the SU(2) gauge transformation in Ref.~\onlinecite{LeeWen06} is that it leaves the spin operator invariant. So (\ref{con3}) is invariant under $SU(2)_1 \times SU(2)_2$. Finally, we note that the $SU(2)_S$ spin rotation in (\ref{eq:gaugeS}) performs an adjoint rotation of (\ref{con3}) in spin space, and so the vanishing of the total spin per site is also obeyed after the $SU(2)_S$ transformation.

\subsection*{Acknowledgements}
%********************************************************

This research was supported by the National Science Foundation under Grant No. DMR-1664842.

\bibliography{pseudogap}

\begin{thebibliography}{41}%
\makeatletter
\providecommand \@ifxundefined [1]{%
 \@ifx{#1\undefined}
}%
\providecommand \@ifnum [1]{%
 \ifnum #1\expandafter \@firstoftwo
 \else \expandafter \@secondoftwo
 \fi
}%
\providecommand \@ifx [1]{%
 \ifx #1\expandafter \@firstoftwo
 \else \expandafter \@secondoftwo
 \fi
}%
\providecommand \natexlab [1]{#1}%
\providecommand \enquote  [1]{``#1''}%
\providecommand \bibnamefont  [1]{#1}%
\providecommand \bibfnamefont [1]{#1}%
\providecommand \citenamefont [1]{#1}%
\providecommand \href@noop [0]{\@secondoftwo}%
\providecommand \href [0]{\begingroup \@sanitize@url \@href}%
\providecommand \@href[1]{\@@startlink{#1}\@@href}%
\providecommand \@@href[1]{\endgroup#1\@@endlink}%
\providecommand \@sanitize@url [0]{\catcode `\\12\catcode `\$12\catcode
  `\&12\catcode `\#12\catcode `\^12\catcode `\_12\catcode `\%12\relax}%
\providecommand \@@startlink[1]{}%
\providecommand \@@endlink[0]{}%
\providecommand \url  [0]{\begingroup\@sanitize@url \@url }%
\providecommand \@url [1]{\endgroup\@href {#1}{\urlprefix }}%
\providecommand \urlprefix  [0]{URL }%
\providecommand \Eprint [0]{\href }%
\providecommand \doibase [0]{http://dx.doi.org/}%
\providecommand \selectlanguage [0]{\@gobble}%
\providecommand \bibinfo  [0]{\@secondoftwo}%
\providecommand \bibfield  [0]{\@secondoftwo}%
\providecommand \translation [1]{[#1]}%
\providecommand \BibitemOpen [0]{}%
\providecommand \bibitemStop [0]{}%
\providecommand \bibitemNoStop [0]{.\EOS\space}%
\providecommand \EOS [0]{\spacefactor3000\relax}%
\providecommand \BibitemShut  [1]{\csname bibitem#1\endcsname}%
\let\auto@bib@innerbib\@empty
%</preamble>
\bibitem [{\citenamefont {{Lee}}\ \emph {et~al.}(2006)\citenamefont {{Lee}},
  \citenamefont {{Nagaosa}},\ and\ \citenamefont {{Wen}}}]{LeeWen06}%
  \BibitemOpen
  \bibfield  {author} {\bibinfo {author} {\bibfnamefont {P.~A.}\ \bibnamefont
  {{Lee}}}, \bibinfo {author} {\bibfnamefont {N.}~\bibnamefont {{Nagaosa}}}, \
  and\ \bibinfo {author} {\bibfnamefont {X.-G.}\ \bibnamefont {{Wen}}},\
  }\bibfield  {title} {\enquote {\bibinfo {title} {{Doping a Mott Insulator:
  Physics of High Temperature Superconductivity}},}\ }\href {\doibase
  10.1103/RevModPhys.78.17} {\bibfield  {journal} {\bibinfo  {journal} {Rev.
  Mod. Phys.}\ }\textbf {\bibinfo {volume} {78}},\ \bibinfo {pages} {17}
  (\bibinfo {year} {2006})},\ \Eprint {http://arxiv.org/abs/cond-mat/0410445}
  {arXiv:cond-mat/0410445 [cond-mat.str-el]} \BibitemShut {NoStop}%
\bibitem [{\citenamefont {{Proust}}\ and\ \citenamefont
  {{Taillefer}}(2019)}]{CPLT18}%
  \BibitemOpen
  \bibfield  {author} {\bibinfo {author} {\bibfnamefont {C.}~\bibnamefont
  {{Proust}}}\ and\ \bibinfo {author} {\bibfnamefont {L.}~\bibnamefont
  {{Taillefer}}},\ }\bibfield  {title} {\enquote {\bibinfo {title} {{The
  Remarkable Underlying Ground States of Cuprate Superconductors}},}\ }\href
  {\doibase 10.1146/annurev-conmatphys-031218-013210} {\bibfield  {journal}
  {\bibinfo  {journal} {Annual Review of Condensed Matter Physics}\ }\textbf
  {\bibinfo {volume} {10}},\ \bibinfo {pages} {409} (\bibinfo {year} {2019})},\
  \Eprint {http://arxiv.org/abs/1807.05074} {arXiv:1807.05074
  [cond-mat.supr-con]} \BibitemShut {NoStop}%
\bibitem [{\citenamefont {{Vishik}}\ \emph {et~al.}(2012)\citenamefont
  {{Vishik}}, \citenamefont {{Hashimoto}}, \citenamefont {{He}}, \citenamefont
  {{Lee}}, \citenamefont {{Schmitt}}, \citenamefont {{Lu}}, \citenamefont
  {{Moore}}, \citenamefont {{Zhang}}, \citenamefont {{Meevasana}},
  \citenamefont {{Sasagawa}}, \citenamefont {{Uchida}}, \citenamefont
  {{Fujita}}, \citenamefont {{Ishida}}, \citenamefont {{Ishikado}},
  \citenamefont {{Yoshida}}, \citenamefont {{Eisaki}}, \citenamefont
  {{Hussain}}, \citenamefont {{Devereaux}},\ and\ \citenamefont
  {{Shen}}}]{Vishik2012}%
  \BibitemOpen
  \bibfield  {author} {\bibinfo {author} {\bibfnamefont {I.~M.}\ \bibnamefont
  {{Vishik}}}, \bibinfo {author} {\bibfnamefont {M.}~\bibnamefont
  {{Hashimoto}}}, \bibinfo {author} {\bibfnamefont {R.-H.}\ \bibnamefont
  {{He}}}, \bibinfo {author} {\bibfnamefont {W.-S.}\ \bibnamefont {{Lee}}},
  \bibinfo {author} {\bibfnamefont {F.}~\bibnamefont {{Schmitt}}}, \bibinfo
  {author} {\bibfnamefont {D.}~\bibnamefont {{Lu}}}, \bibinfo {author}
  {\bibfnamefont {R.~G.}\ \bibnamefont {{Moore}}}, \bibinfo {author}
  {\bibfnamefont {C.}~\bibnamefont {{Zhang}}}, \bibinfo {author} {\bibfnamefont
  {W.}~\bibnamefont {{Meevasana}}}, \bibinfo {author} {\bibfnamefont
  {T.}~\bibnamefont {{Sasagawa}}}, \bibinfo {author} {\bibfnamefont
  {S.}~\bibnamefont {{Uchida}}}, \bibinfo {author} {\bibfnamefont
  {K.}~\bibnamefont {{Fujita}}}, \bibinfo {author} {\bibfnamefont
  {S.}~\bibnamefont {{Ishida}}}, \bibinfo {author} {\bibfnamefont
  {M.}~\bibnamefont {{Ishikado}}}, \bibinfo {author} {\bibfnamefont
  {Y.}~\bibnamefont {{Yoshida}}}, \bibinfo {author} {\bibfnamefont
  {H.}~\bibnamefont {{Eisaki}}}, \bibinfo {author} {\bibfnamefont
  {Z.}~\bibnamefont {{Hussain}}}, \bibinfo {author} {\bibfnamefont {T.~P.}\
  \bibnamefont {{Devereaux}}}, \ and\ \bibinfo {author} {\bibfnamefont {Z.-X.}\
  \bibnamefont {{Shen}}},\ }\bibfield  {title} {\enquote {\bibinfo {title}
  {{Phase competition in trisected superconducting dome}},}\ }\href {\doibase
  10.1073/pnas.1209471109} {\bibfield  {journal} {\bibinfo  {journal}
  {Proceedings of the National Academy of Science}\ }\textbf {\bibinfo {volume}
  {109}},\ \bibinfo {pages} {18332} (\bibinfo {year} {2012})},\ \Eprint
  {http://arxiv.org/abs/1209.6514} {arXiv:1209.6514 [cond-mat.supr-con]}
  \BibitemShut {NoStop}%
\bibitem [{\citenamefont {{He}}\ \emph {et~al.}(2014)\citenamefont {{He}},
  \citenamefont {{Yin}}, \citenamefont {{Zech}}, \citenamefont
  {{Soumyanarayanan}}, \citenamefont {{Yee}}, \citenamefont {{Williams}},
  \citenamefont {{Boyer}}, \citenamefont {{Chatterjee}}, \citenamefont
  {{Wise}}, \citenamefont {{Zeljkovic}}, \citenamefont {{Kondo}}, \citenamefont
  {{Takeuchi}}, \citenamefont {{Ikuta}}, \citenamefont {{Mistark}},
  \citenamefont {{Markiewicz}}, \citenamefont {{Bansil}}, \citenamefont
  {{Sachdev}}, \citenamefont {{Hudson}},\ and\ \citenamefont
  {{Hoffman}}}]{He14}%
  \BibitemOpen
  \bibfield  {author} {\bibinfo {author} {\bibfnamefont {Y.}~\bibnamefont
  {{He}}}, \bibinfo {author} {\bibfnamefont {Y.}~\bibnamefont {{Yin}}},
  \bibinfo {author} {\bibfnamefont {M.}~\bibnamefont {{Zech}}}, \bibinfo
  {author} {\bibfnamefont {A.}~\bibnamefont {{Soumyanarayanan}}}, \bibinfo
  {author} {\bibfnamefont {M.~M.}\ \bibnamefont {{Yee}}}, \bibinfo {author}
  {\bibfnamefont {T.}~\bibnamefont {{Williams}}}, \bibinfo {author}
  {\bibfnamefont {M.~C.}\ \bibnamefont {{Boyer}}}, \bibinfo {author}
  {\bibfnamefont {K.}~\bibnamefont {{Chatterjee}}}, \bibinfo {author}
  {\bibfnamefont {W.~D.}\ \bibnamefont {{Wise}}}, \bibinfo {author}
  {\bibfnamefont {I.}~\bibnamefont {{Zeljkovic}}}, \bibinfo {author}
  {\bibfnamefont {T.}~\bibnamefont {{Kondo}}}, \bibinfo {author} {\bibfnamefont
  {T.}~\bibnamefont {{Takeuchi}}}, \bibinfo {author} {\bibfnamefont
  {H.}~\bibnamefont {{Ikuta}}}, \bibinfo {author} {\bibfnamefont
  {P.}~\bibnamefont {{Mistark}}}, \bibinfo {author} {\bibfnamefont {R.~S.}\
  \bibnamefont {{Markiewicz}}}, \bibinfo {author} {\bibfnamefont
  {A.}~\bibnamefont {{Bansil}}}, \bibinfo {author} {\bibfnamefont
  {S.}~\bibnamefont {{Sachdev}}}, \bibinfo {author} {\bibfnamefont {E.~W.}\
  \bibnamefont {{Hudson}}}, \ and\ \bibinfo {author} {\bibfnamefont {J.~E.}\
  \bibnamefont {{Hoffman}}},\ }\bibfield  {title} {\enquote {\bibinfo {title}
  {{Fermi Surface and Pseudogap Evolution in a Cuprate Superconductor}},}\
  }\href {\doibase 10.1126/science.1248221} {\bibfield  {journal} {\bibinfo
  {journal} {Science}\ }\textbf {\bibinfo {volume} {344}},\ \bibinfo {pages}
  {608} (\bibinfo {year} {2014})},\ \Eprint {http://arxiv.org/abs/1305.2778}
  {arXiv:1305.2778 [cond-mat.supr-con]} \BibitemShut {NoStop}%
\bibitem [{\citenamefont {{Fujita}}\ \emph {et~al.}(2014)\citenamefont
  {{Fujita}}, \citenamefont {{Kim}}, \citenamefont {{Lee}}, \citenamefont
  {{Lee}}, \citenamefont {{Hamidian}}, \citenamefont {{Firmo}}, \citenamefont
  {{Mukhopadhyay}}, \citenamefont {{Eisaki}}, \citenamefont {{Uchida}},
  \citenamefont {{Lawler}}, \citenamefont {{Kim}},\ and\ \citenamefont
  {{Davis}}}]{Fujita14}%
  \BibitemOpen
  \bibfield  {author} {\bibinfo {author} {\bibfnamefont {K.}~\bibnamefont
  {{Fujita}}}, \bibinfo {author} {\bibfnamefont {C.~K.}\ \bibnamefont {{Kim}}},
  \bibinfo {author} {\bibfnamefont {I.}~\bibnamefont {{Lee}}}, \bibinfo
  {author} {\bibfnamefont {J.}~\bibnamefont {{Lee}}}, \bibinfo {author}
  {\bibfnamefont {M.~H.}\ \bibnamefont {{Hamidian}}}, \bibinfo {author}
  {\bibfnamefont {I.~A.}\ \bibnamefont {{Firmo}}}, \bibinfo {author}
  {\bibfnamefont {S.}~\bibnamefont {{Mukhopadhyay}}}, \bibinfo {author}
  {\bibfnamefont {H.}~\bibnamefont {{Eisaki}}}, \bibinfo {author}
  {\bibfnamefont {S.}~\bibnamefont {{Uchida}}}, \bibinfo {author}
  {\bibfnamefont {M.~J.}\ \bibnamefont {{Lawler}}}, \bibinfo {author}
  {\bibfnamefont {E.-A.}\ \bibnamefont {{Kim}}}, \ and\ \bibinfo {author}
  {\bibfnamefont {J.~C.}\ \bibnamefont {{Davis}}},\ }\bibfield  {title}
  {\enquote {\bibinfo {title} {{Simultaneous Transitions in Cuprate
  Momentum-Space Topology and Electronic Symmetry Breaking}},}\ }\href
  {\doibase 10.1126/science.1248783} {\bibfield  {journal} {\bibinfo  {journal}
  {Science}\ }\textbf {\bibinfo {volume} {344}},\ \bibinfo {pages} {612}
  (\bibinfo {year} {2014})},\ \Eprint {http://arxiv.org/abs/1403.7788}
  {arXiv:1403.7788 [cond-mat.supr-con]} \BibitemShut {NoStop}%
\bibitem [{\citenamefont {{Badoux}}\ \emph {et~al.}(2016)\citenamefont
  {{Badoux}}, \citenamefont {{Tabis}}, \citenamefont {{Lalibert{\'e}}},
  \citenamefont {{Grissonnanche}}, \citenamefont {{Vignolle}}, \citenamefont
  {{Vignolles}}, \citenamefont {{B{\'e}ard}}, \citenamefont {{Bonn}},
  \citenamefont {{Hardy}}, \citenamefont {{Liang}}, \citenamefont
  {{Doiron-Leyraud}}, \citenamefont {{Taillefer}},\ and\ \citenamefont
  {{Proust}}}]{Badoux16}%
  \BibitemOpen
  \bibfield  {author} {\bibinfo {author} {\bibfnamefont {S.}~\bibnamefont
  {{Badoux}}}, \bibinfo {author} {\bibfnamefont {W.}~\bibnamefont {{Tabis}}},
  \bibinfo {author} {\bibfnamefont {F.}~\bibnamefont {{Lalibert{\'e}}}},
  \bibinfo {author} {\bibfnamefont {G.}~\bibnamefont {{Grissonnanche}}},
  \bibinfo {author} {\bibfnamefont {B.}~\bibnamefont {{Vignolle}}}, \bibinfo
  {author} {\bibfnamefont {D.}~\bibnamefont {{Vignolles}}}, \bibinfo {author}
  {\bibfnamefont {J.}~\bibnamefont {{B{\'e}ard}}}, \bibinfo {author}
  {\bibfnamefont {D.~A.}\ \bibnamefont {{Bonn}}}, \bibinfo {author}
  {\bibfnamefont {W.~N.}\ \bibnamefont {{Hardy}}}, \bibinfo {author}
  {\bibfnamefont {R.}~\bibnamefont {{Liang}}}, \bibinfo {author} {\bibfnamefont
  {N.}~\bibnamefont {{Doiron-Leyraud}}}, \bibinfo {author} {\bibfnamefont
  {L.}~\bibnamefont {{Taillefer}}}, \ and\ \bibinfo {author} {\bibfnamefont
  {C.}~\bibnamefont {{Proust}}},\ }\bibfield  {title} {\enquote {\bibinfo
  {title} {{Change of carrier density at the pseudogap critical point of a
  cuprate superconductor}},}\ }\href {\doibase 10.1038/nature16983} {\bibfield
  {journal} {\bibinfo  {journal} {Nature}\ }\textbf {\bibinfo {volume} {531}},\
  \bibinfo {pages} {210} (\bibinfo {year} {2016})},\ \Eprint
  {http://arxiv.org/abs/1511.08162} {arXiv:1511.08162 [cond-mat.supr-con]}
  \BibitemShut {NoStop}%
\bibitem [{\citenamefont {Loram}\ \emph {et~al.}(2001)\citenamefont {Loram},
  \citenamefont {Luo}, \citenamefont {Cooper}, \citenamefont {Liang},\ and\
  \citenamefont {Tallon}}]{Loram01}%
  \BibitemOpen
  \bibfield  {author} {\bibinfo {author} {\bibfnamefont {J.}~\bibnamefont
  {Loram}}, \bibinfo {author} {\bibfnamefont {J.}~\bibnamefont {Luo}}, \bibinfo
  {author} {\bibfnamefont {J.}~\bibnamefont {Cooper}}, \bibinfo {author}
  {\bibfnamefont {W.}~\bibnamefont {Liang}}, \ and\ \bibinfo {author}
  {\bibfnamefont {J.}~\bibnamefont {Tallon}},\ }\bibfield  {title} {\enquote
  {\bibinfo {title} {Evidence on the pseudogap and condensate from the
  electronic specific heat},}\ }\href {\doibase
  https://doi.org/10.1016/S0022-3697(00)00101-3} {\bibfield  {journal}
  {\bibinfo  {journal} {Journal of Physics and Chemistry of Solids}\ }\textbf
  {\bibinfo {volume} {62}},\ \bibinfo {pages} {59 } (\bibinfo {year}
  {2001})}\BibitemShut {NoStop}%
\bibitem [{\citenamefont {{Tallon}}\ \emph {et~al.}(2019)\citenamefont
  {{Tallon}}, \citenamefont {{Storey}}, \citenamefont {{Cooper}},\ and\
  \citenamefont {{Loram}}}]{Loram19}%
  \BibitemOpen
  \bibfield  {author} {\bibinfo {author} {\bibfnamefont {J.~L.}\ \bibnamefont
  {{Tallon}}}, \bibinfo {author} {\bibfnamefont {J.~G.}\ \bibnamefont
  {{Storey}}}, \bibinfo {author} {\bibfnamefont {J.~R.}\ \bibnamefont
  {{Cooper}}}, \ and\ \bibinfo {author} {\bibfnamefont {J.~W.}\ \bibnamefont
  {{Loram}}},\ }\bibfield  {title} {\enquote {\bibinfo {title} {{Locating the
  pseudogap closing point in cuprate superconductors: absence of entrant or
  reentrant behavior}},}\ }\href@noop {} {\  (\bibinfo {year} {2019})},\
  \Eprint {http://arxiv.org/abs/1907.12018} {arXiv:1907.12018
  [cond-mat.supr-con]} \BibitemShut {NoStop}%
\bibitem [{\citenamefont {{Michon}}\ \emph {et~al.}(2019)\citenamefont
  {{Michon}}, \citenamefont {{Girod}}, \citenamefont {{Badoux}}, \citenamefont
  {{Ka{\v{c}}mar{\v{c}}{\'\i}k}}, \citenamefont {{Ma}}, \citenamefont
  {{Dragomir}}, \citenamefont {{Dabkowska}}, \citenamefont {{Gaulin}},
  \citenamefont {{Zhou}}, \citenamefont {{Pyon}}, \citenamefont {{Takayama}},
  \citenamefont {{Takagi}}, \citenamefont {{Verret}}, \citenamefont
  {{Doiron-Leyraud}}, \citenamefont {{Marcenat}}, \citenamefont {{Taillefer}},\
  and\ \citenamefont {{Klein}}}]{Michon18}%
  \BibitemOpen
  \bibfield  {author} {\bibinfo {author} {\bibfnamefont {B.}~\bibnamefont
  {{Michon}}}, \bibinfo {author} {\bibfnamefont {C.}~\bibnamefont {{Girod}}},
  \bibinfo {author} {\bibfnamefont {S.}~\bibnamefont {{Badoux}}}, \bibinfo
  {author} {\bibfnamefont {J.}~\bibnamefont {{Ka{\v{c}}mar{\v{c}}{\'\i}k}}},
  \bibinfo {author} {\bibfnamefont {Q.}~\bibnamefont {{Ma}}}, \bibinfo {author}
  {\bibfnamefont {M.}~\bibnamefont {{Dragomir}}}, \bibinfo {author}
  {\bibfnamefont {H.~A.}\ \bibnamefont {{Dabkowska}}}, \bibinfo {author}
  {\bibfnamefont {B.~D.}\ \bibnamefont {{Gaulin}}}, \bibinfo {author}
  {\bibfnamefont {J.~S.}\ \bibnamefont {{Zhou}}}, \bibinfo {author}
  {\bibfnamefont {S.}~\bibnamefont {{Pyon}}}, \bibinfo {author} {\bibfnamefont
  {T.}~\bibnamefont {{Takayama}}}, \bibinfo {author} {\bibfnamefont
  {H.}~\bibnamefont {{Takagi}}}, \bibinfo {author} {\bibfnamefont
  {S.}~\bibnamefont {{Verret}}}, \bibinfo {author} {\bibfnamefont
  {N.}~\bibnamefont {{Doiron-Leyraud}}}, \bibinfo {author} {\bibfnamefont
  {C.}~\bibnamefont {{Marcenat}}}, \bibinfo {author} {\bibfnamefont
  {L.}~\bibnamefont {{Taillefer}}}, \ and\ \bibinfo {author} {\bibfnamefont
  {T.}~\bibnamefont {{Klein}}},\ }\bibfield  {title} {\enquote {\bibinfo
  {title} {{Thermodynamic signatures of quantum criticality in cuprate
  superconductors}},}\ }\href {\doibase 10.1038/s41586-019-0932-x} {\bibfield
  {journal} {\bibinfo  {journal} {Nature}\ }\textbf {\bibinfo {volume} {567}},\
  \bibinfo {pages} {218} (\bibinfo {year} {2019})},\ \Eprint
  {http://arxiv.org/abs/1804.08502} {arXiv:1804.08502 [cond-mat.supr-con]}
  \BibitemShut {NoStop}%
\bibitem [{\citenamefont {{Tang}}\ \emph {et~al.}(2018)\citenamefont {{Tang}},
  \citenamefont {{Mangin-Thro}}, \citenamefont {{Wildes}}, \citenamefont
  {{Chan}}, \citenamefont {{Dorow}}, \citenamefont {{Jeong}}, \citenamefont
  {{Sidis}}, \citenamefont {{Greven}},\ and\ \citenamefont
  {{Bourges}}}]{Bourges18}%
  \BibitemOpen
  \bibfield  {author} {\bibinfo {author} {\bibfnamefont {Y.}~\bibnamefont
  {{Tang}}}, \bibinfo {author} {\bibfnamefont {L.}~\bibnamefont
  {{Mangin-Thro}}}, \bibinfo {author} {\bibfnamefont {A.}~\bibnamefont
  {{Wildes}}}, \bibinfo {author} {\bibfnamefont {M.~K.}\ \bibnamefont
  {{Chan}}}, \bibinfo {author} {\bibfnamefont {C.~J.}\ \bibnamefont {{Dorow}}},
  \bibinfo {author} {\bibfnamefont {J.}~\bibnamefont {{Jeong}}}, \bibinfo
  {author} {\bibfnamefont {Y.}~\bibnamefont {{Sidis}}}, \bibinfo {author}
  {\bibfnamefont {M.}~\bibnamefont {{Greven}}}, \ and\ \bibinfo {author}
  {\bibfnamefont {P.}~\bibnamefont {{Bourges}}},\ }\bibfield  {title} {\enquote
  {\bibinfo {title} {{Orientation of the intra-unit-cell magnetic moment in the
  high-$T_{c}$ superconductor HgBa$_{2}$CuO$_{4 +{\ensuremath{\delta}}}$}},}\
  }\href {\doibase 10.1103/PhysRevB.98.214418} {\bibfield  {journal} {\bibinfo
  {journal} {\prb}\ }\textbf {\bibinfo {volume} {98}},\ \bibinfo {eid} {214418}
  (\bibinfo {year} {2018})},\ \Eprint {http://arxiv.org/abs/1805.02063}
  {arXiv:1805.02063 [cond-mat.supr-con]} \BibitemShut {NoStop}%
\bibitem [{\citenamefont {Chen}\ \emph {et~al.}(2019)\citenamefont {Chen},
  \citenamefont {Hashimoto}, \citenamefont {He}, \citenamefont {Song},
  \citenamefont {Xu}, \citenamefont {He}, \citenamefont {Devereaux},
  \citenamefont {Eisaki}, \citenamefont {Lu}, \citenamefont {Zaanen},\ and\
  \citenamefont {Shen}}]{Shen19}%
  \BibitemOpen
  \bibfield  {author} {\bibinfo {author} {\bibfnamefont {S.-D.}\ \bibnamefont
  {Chen}}, \bibinfo {author} {\bibfnamefont {M.}~\bibnamefont {Hashimoto}},
  \bibinfo {author} {\bibfnamefont {Y.}~\bibnamefont {He}}, \bibinfo {author}
  {\bibfnamefont {D.}~\bibnamefont {Song}}, \bibinfo {author} {\bibfnamefont
  {K.-J.}\ \bibnamefont {Xu}}, \bibinfo {author} {\bibfnamefont {J.-F.}\
  \bibnamefont {He}}, \bibinfo {author} {\bibfnamefont {T.~P.}\ \bibnamefont
  {Devereaux}}, \bibinfo {author} {\bibfnamefont {H.}~\bibnamefont {Eisaki}},
  \bibinfo {author} {\bibfnamefont {D.-H.}\ \bibnamefont {Lu}}, \bibinfo
  {author} {\bibfnamefont {J.}~\bibnamefont {Zaanen}}, \ and\ \bibinfo {author}
  {\bibfnamefont {Z.-X.}\ \bibnamefont {Shen}},\ }\bibfield  {title} {\enquote
  {\bibinfo {title} {{Incoherent strange metal sharply bounded by a critical
  doping in Bi2212}},}\ }\href {\doibase 10.1126/science.aaw8850} {\bibfield
  {journal} {\bibinfo  {journal} {Science}\ }\textbf {\bibinfo {volume}
  {366}},\ \bibinfo {pages} {1099} (\bibinfo {year} {2019})}\BibitemShut
  {NoStop}%
\bibitem [{\citenamefont {{Panagopoulos}}\ \emph {et~al.}(2002)\citenamefont
  {{Panagopoulos}}, \citenamefont {{Tallon}}, \citenamefont {{Rainford}},
  \citenamefont {{Xiang}}, \citenamefont {{Cooper}},\ and\ \citenamefont
  {{Scott}}}]{CPana1}%
  \BibitemOpen
  \bibfield  {author} {\bibinfo {author} {\bibfnamefont {C.}~\bibnamefont
  {{Panagopoulos}}}, \bibinfo {author} {\bibfnamefont {J.~L.}\ \bibnamefont
  {{Tallon}}}, \bibinfo {author} {\bibfnamefont {B.~D.}\ \bibnamefont
  {{Rainford}}}, \bibinfo {author} {\bibfnamefont {T.}~\bibnamefont {{Xiang}}},
  \bibinfo {author} {\bibfnamefont {J.~R.}\ \bibnamefont {{Cooper}}}, \ and\
  \bibinfo {author} {\bibfnamefont {C.~A.}\ \bibnamefont {{Scott}}},\
  }\bibfield  {title} {\enquote {\bibinfo {title} {{Evidence for a generic
  quantum transition in high-$T_{c}$ cuprates}},}\ }\href {\doibase
  10.1103/PhysRevB.66.064501} {\bibfield  {journal} {\bibinfo  {journal} {Phys.
  Rev. B}\ }\textbf {\bibinfo {volume} {66}},\ \bibinfo {eid} {064501}
  (\bibinfo {year} {2002})},\ \Eprint {http://arxiv.org/abs/cond-mat/0204106}
  {arXiv:cond-mat/0204106 [cond-mat.supr-con]} \BibitemShut {NoStop}%
\bibitem [{\citenamefont {{Panagopoulos}}\ \emph {et~al.}(2004)\citenamefont
  {{Panagopoulos}}, \citenamefont {{Petrovic}}, \citenamefont {{Hillier}},
  \citenamefont {{Tallon}}, \citenamefont {{Scott}},\ and\ \citenamefont
  {{Rainford}}}]{CPana2}%
  \BibitemOpen
  \bibfield  {author} {\bibinfo {author} {\bibfnamefont {C.}~\bibnamefont
  {{Panagopoulos}}}, \bibinfo {author} {\bibfnamefont {A.~P.}\ \bibnamefont
  {{Petrovic}}}, \bibinfo {author} {\bibfnamefont {A.~D.}\ \bibnamefont
  {{Hillier}}}, \bibinfo {author} {\bibfnamefont {J.~L.}\ \bibnamefont
  {{Tallon}}}, \bibinfo {author} {\bibfnamefont {C.~A.}\ \bibnamefont
  {{Scott}}}, \ and\ \bibinfo {author} {\bibfnamefont {B.~D.}\ \bibnamefont
  {{Rainford}}},\ }\bibfield  {title} {\enquote {\bibinfo {title} {{Exposing
  the spin-glass ground state of the nonsuperconducting
  La$_{2-x}$Sr$_{x}$Cu$_{1-y}$Zn$_{y}$O$_{4}$ high-$T_{c}$ oxide}},}\ }\href
  {\doibase 10.1103/PhysRevB.69.144510} {\bibfield  {journal} {\bibinfo
  {journal} {Phys. Rev. B}\ }\textbf {\bibinfo {volume} {69}},\ \bibinfo {eid}
  {144510} (\bibinfo {year} {2004})},\ \Eprint
  {http://arxiv.org/abs/cond-mat/0307392} {arXiv:cond-mat/0307392
  [cond-mat.supr-con]} \BibitemShut {NoStop}%
\bibitem [{\citenamefont {{Frachet}}\ \emph {et~al.}(2019)\citenamefont
  {{Frachet}}, \citenamefont {{Vinograd}}, \citenamefont {{Zhou}},
  \citenamefont {{Benhabib}}, \citenamefont {{Wu}}, \citenamefont {{Mayaffre}},
  \citenamefont {{Kr{\"a}mer}}, \citenamefont {{Ramakrishna}}, \citenamefont
  {{Reyes}}, \citenamefont {{Debray}}, \citenamefont {{Kurosawa}},
  \citenamefont {{Momono}}, \citenamefont {{Oda}}, \citenamefont {{Komiya}},
  \citenamefont {{Ono}}, \citenamefont {{Horio}}, \citenamefont {{Chang}},
  \citenamefont {{Proust}}, \citenamefont {{LeBoeuf}},\ and\ \citenamefont
  {{Julien}}}]{Julien19}%
  \BibitemOpen
  \bibfield  {author} {\bibinfo {author} {\bibfnamefont {M.}~\bibnamefont
  {{Frachet}}}, \bibinfo {author} {\bibfnamefont {I.}~\bibnamefont
  {{Vinograd}}}, \bibinfo {author} {\bibfnamefont {R.}~\bibnamefont {{Zhou}}},
  \bibinfo {author} {\bibfnamefont {S.}~\bibnamefont {{Benhabib}}}, \bibinfo
  {author} {\bibfnamefont {S.}~\bibnamefont {{Wu}}}, \bibinfo {author}
  {\bibfnamefont {H.}~\bibnamefont {{Mayaffre}}}, \bibinfo {author}
  {\bibfnamefont {S.}~\bibnamefont {{Kr{\"a}mer}}}, \bibinfo {author}
  {\bibfnamefont {S.~K.}\ \bibnamefont {{Ramakrishna}}}, \bibinfo {author}
  {\bibfnamefont {A.}~\bibnamefont {{Reyes}}}, \bibinfo {author} {\bibfnamefont
  {J.}~\bibnamefont {{Debray}}}, \bibinfo {author} {\bibfnamefont
  {T.}~\bibnamefont {{Kurosawa}}}, \bibinfo {author} {\bibfnamefont
  {N.}~\bibnamefont {{Momono}}}, \bibinfo {author} {\bibfnamefont
  {M.}~\bibnamefont {{Oda}}}, \bibinfo {author} {\bibfnamefont
  {S.}~\bibnamefont {{Komiya}}}, \bibinfo {author} {\bibfnamefont
  {S.}~\bibnamefont {{Ono}}}, \bibinfo {author} {\bibfnamefont
  {M.}~\bibnamefont {{Horio}}}, \bibinfo {author} {\bibfnamefont
  {J.}~\bibnamefont {{Chang}}}, \bibinfo {author} {\bibfnamefont
  {C.}~\bibnamefont {{Proust}}}, \bibinfo {author} {\bibfnamefont
  {D.}~\bibnamefont {{LeBoeuf}}}, \ and\ \bibinfo {author} {\bibfnamefont
  {M.-H.}\ \bibnamefont {{Julien}}},\ }\bibfield  {title} {\enquote {\bibinfo
  {title} {{Hidden magnetism at the pseudogap critical point of a high
  temperature superconductor}},}\ }\href@noop {} {\  (\bibinfo {year}
  {2019})},\ \Eprint {http://arxiv.org/abs/1909.10258} {arXiv:1909.10258
  [cond-mat.supr-con]} \BibitemShut {NoStop}%
\bibitem [{\citenamefont {{Senthil}}\ \emph {et~al.}(2003)\citenamefont
  {{Senthil}}, \citenamefont {{Sachdev}},\ and\ \citenamefont
  {{Vojta}}}]{Senthil_2003}%
  \BibitemOpen
  \bibfield  {author} {\bibinfo {author} {\bibfnamefont {T.}~\bibnamefont
  {{Senthil}}}, \bibinfo {author} {\bibfnamefont {S.}~\bibnamefont
  {{Sachdev}}}, \ and\ \bibinfo {author} {\bibfnamefont {M.}~\bibnamefont
  {{Vojta}}},\ }\bibfield  {title} {\enquote {\bibinfo {title} {{Fractionalized
  Fermi Liquids}},}\ }\href {\doibase 10.1103/PhysRevLett.90.216403} {\bibfield
   {journal} {\bibinfo  {journal} {Phys. Rev. Lett.}\ }\textbf {\bibinfo
  {volume} {90}},\ \bibinfo {eid} {216403} (\bibinfo {year} {2003})},\ \Eprint
  {http://arxiv.org/abs/cond-mat/0209144} {cond-mat/0209144} \BibitemShut
  {NoStop}%
\bibitem [{\citenamefont {{Senthil}}\ \emph {et~al.}(2004)\citenamefont
  {{Senthil}}, \citenamefont {{Vojta}},\ and\ \citenamefont
  {{Sachdev}}}]{SVS04}%
  \BibitemOpen
  \bibfield  {author} {\bibinfo {author} {\bibfnamefont {T.}~\bibnamefont
  {{Senthil}}}, \bibinfo {author} {\bibfnamefont {M.}~\bibnamefont {{Vojta}}},
  \ and\ \bibinfo {author} {\bibfnamefont {S.}~\bibnamefont {{Sachdev}}},\
  }\bibfield  {title} {\enquote {\bibinfo {title} {{Weak magnetism and
  non-Fermi liquids near heavy-fermion critical points}},}\ }\href {\doibase
  10.1103/PhysRevB.69.035111} {\bibfield  {journal} {\bibinfo  {journal} {Phys.
  Rev. B}\ }\textbf {\bibinfo {volume} {69}},\ \bibinfo {eid} {035111}
  (\bibinfo {year} {2004})},\ \Eprint {http://arxiv.org/abs/cond-mat/0305193}
  {arXiv:cond-mat/0305193 [cond-mat.str-el]} \BibitemShut {NoStop}%
\bibitem [{\citenamefont {{Paramekanti}}\ and\ \citenamefont
  {{Vishwanath}}(2004)}]{Paramekanti_2004}%
  \BibitemOpen
  \bibfield  {author} {\bibinfo {author} {\bibfnamefont {A.}~\bibnamefont
  {{Paramekanti}}}\ and\ \bibinfo {author} {\bibfnamefont {A.}~\bibnamefont
  {{Vishwanath}}},\ }\bibfield  {title} {\enquote {\bibinfo {title} {{Extending
  Luttinger's theorem to $\mathbb{Z}_{2}$ fractionalized phases of matter}},}\
  }\href {\doibase 10.1103/PhysRevB.70.245118} {\bibfield  {journal} {\bibinfo
  {journal} {Phys. Rev. B}\ }\textbf {\bibinfo {volume} {70}},\ \bibinfo {eid}
  {245118} (\bibinfo {year} {2004})},\ \Eprint
  {http://arxiv.org/abs/cond-mat/0406619} {arXiv:cond-mat/0406619
  [cond-mat.str-el]} \BibitemShut {NoStop}%
\bibitem [{\citenamefont {{Wen}}\ and\ \citenamefont {{Lee}}(1996)}]{WenLee96}%
  \BibitemOpen
  \bibfield  {author} {\bibinfo {author} {\bibfnamefont {X.-G.}\ \bibnamefont
  {{Wen}}}\ and\ \bibinfo {author} {\bibfnamefont {P.~A.}\ \bibnamefont
  {{Lee}}},\ }\bibfield  {title} {\enquote {\bibinfo {title} {{Theory of
  Underdoped Cuprates}},}\ }\href {\doibase 10.1103/PhysRevLett.76.503}
  {\bibfield  {journal} {\bibinfo  {journal} {Phys. Rev. Lett.}\ }\textbf
  {\bibinfo {volume} {76}},\ \bibinfo {pages} {503} (\bibinfo {year} {1996})},\
  \Eprint {http://arxiv.org/abs/cond-mat/9506065} {arXiv:cond-mat/9506065
  [cond-mat]} \BibitemShut {NoStop}%
\bibitem [{\citenamefont {{Yang}}\ \emph {et~al.}(2006)\citenamefont {{Yang}},
  \citenamefont {{Rice}},\ and\ \citenamefont {{Zhang}}}]{YRZ06}%
  \BibitemOpen
  \bibfield  {author} {\bibinfo {author} {\bibfnamefont {K.-Y.}\ \bibnamefont
  {{Yang}}}, \bibinfo {author} {\bibfnamefont {T.~M.}\ \bibnamefont {{Rice}}},
  \ and\ \bibinfo {author} {\bibfnamefont {F.-C.}\ \bibnamefont {{Zhang}}},\
  }\bibfield  {title} {\enquote {\bibinfo {title} {{Phenomenological theory of
  the pseudogap state}},}\ }\href {\doibase 10.1103/PhysRevB.73.174501}
  {\bibfield  {journal} {\bibinfo  {journal} {Phys. Rev. B}\ }\textbf {\bibinfo
  {volume} {73}},\ \bibinfo {eid} {174501} (\bibinfo {year} {2006})},\ \Eprint
  {http://arxiv.org/abs/cond-mat/0602164} {arXiv:cond-mat/0602164
  [cond-mat.supr-con]} \BibitemShut {NoStop}%
\bibitem [{\citenamefont {{Qi}}\ and\ \citenamefont
  {{Sachdev}}(2010)}]{YQSS10}%
  \BibitemOpen
  \bibfield  {author} {\bibinfo {author} {\bibfnamefont {Y.}~\bibnamefont
  {{Qi}}}\ and\ \bibinfo {author} {\bibfnamefont {S.}~\bibnamefont
  {{Sachdev}}},\ }\bibfield  {title} {\enquote {\bibinfo {title} {{Effective
  theory of Fermi pockets in fluctuating antiferromagnets}},}\ }\href {\doibase
  10.1103/PhysRevB.81.115129} {\bibfield  {journal} {\bibinfo  {journal} {Phys.
  Rev. B}\ }\textbf {\bibinfo {volume} {81}},\ \bibinfo {eid} {115129}
  (\bibinfo {year} {2010})},\ \Eprint {http://arxiv.org/abs/0912.0943}
  {arXiv:0912.0943 [cond-mat.str-el]} \BibitemShut {NoStop}%
\bibitem [{\citenamefont {{Mei}}\ \emph {et~al.}(2012)\citenamefont {{Mei}},
  \citenamefont {{Kawasaki}}, \citenamefont {{Zheng}}, \citenamefont {{Weng}},\
  and\ \citenamefont {{Wen}}}]{Mei12}%
  \BibitemOpen
  \bibfield  {author} {\bibinfo {author} {\bibfnamefont {J.-W.}\ \bibnamefont
  {{Mei}}}, \bibinfo {author} {\bibfnamefont {S.}~\bibnamefont {{Kawasaki}}},
  \bibinfo {author} {\bibfnamefont {G.-Q.}\ \bibnamefont {{Zheng}}}, \bibinfo
  {author} {\bibfnamefont {Z.-Y.}\ \bibnamefont {{Weng}}}, \ and\ \bibinfo
  {author} {\bibfnamefont {X.-G.}\ \bibnamefont {{Wen}}},\ }\bibfield  {title}
  {\enquote {\bibinfo {title} {{Luttinger-volume violating Fermi liquid in the
  pseudogap phase of the cuprate superconductors}},}\ }\href {\doibase
  10.1103/PhysRevB.85.134519} {\bibfield  {journal} {\bibinfo  {journal} {Phys.
  Rev. B}\ }\textbf {\bibinfo {volume} {85}},\ \bibinfo {eid} {134519}
  (\bibinfo {year} {2012})},\ \Eprint {http://arxiv.org/abs/1109.0406}
  {arXiv:1109.0406 [cond-mat.supr-con]} \BibitemShut {NoStop}%
\bibitem [{\citenamefont {{Punk}}\ and\ \citenamefont
  {{Sachdev}}(2012)}]{Punk_2012}%
  \BibitemOpen
  \bibfield  {author} {\bibinfo {author} {\bibfnamefont {M.}~\bibnamefont
  {{Punk}}}\ and\ \bibinfo {author} {\bibfnamefont {S.}~\bibnamefont
  {{Sachdev}}},\ }\bibfield  {title} {\enquote {\bibinfo {title} {{Fermi
  surface reconstruction in hole-doped $t$-$J$ models without long-range
  antiferromagnetic order}},}\ }\href {\doibase 10.1103/PhysRevB.85.195123}
  {\bibfield  {journal} {\bibinfo  {journal} {Phys. Rev. B}\ }\textbf {\bibinfo
  {volume} {85}},\ \bibinfo {eid} {195123} (\bibinfo {year} {2012})},\ \Eprint
  {http://arxiv.org/abs/1202.4023} {arXiv:1202.4023 [cond-mat.str-el]}
  \BibitemShut {NoStop}%
\bibitem [{\citenamefont {Punk}\ \emph {et~al.}(2015)\citenamefont {Punk},
  \citenamefont {Allais},\ and\ \citenamefont {Sachdev}}]{Punk_2015}%
  \BibitemOpen
  \bibfield  {author} {\bibinfo {author} {\bibfnamefont {M.}~\bibnamefont
  {Punk}}, \bibinfo {author} {\bibfnamefont {A.}~\bibnamefont {Allais}}, \ and\
  \bibinfo {author} {\bibfnamefont {S.}~\bibnamefont {Sachdev}},\ }\bibfield
  {title} {\enquote {\bibinfo {title} {{A quantum dimer model for the pseudogap
  metal}},}\ }\href {\doibase 10.1073/pnas.1512206112} {\bibfield  {journal}
  {\bibinfo  {journal} {Proc. Nat. Acad. Sci.}\ }\textbf {\bibinfo {volume}
  {112}},\ \bibinfo {pages} {9552} (\bibinfo {year} {2015})},\ \Eprint
  {http://arxiv.org/abs/1501.00978} {arXiv:1501.00978 [cond-mat.str-el]}
  \BibitemShut {NoStop}%
%%CITATION = ARXIV:1501.00978;%%
\bibitem [{\citenamefont {{Feldmeier}}\ \emph {et~al.}(2018)\citenamefont
  {{Feldmeier}}, \citenamefont {{Huber}},\ and\ \citenamefont
  {{Punk}}}]{Punk17}%
  \BibitemOpen
  \bibfield  {author} {\bibinfo {author} {\bibfnamefont {J.}~\bibnamefont
  {{Feldmeier}}}, \bibinfo {author} {\bibfnamefont {S.}~\bibnamefont
  {{Huber}}}, \ and\ \bibinfo {author} {\bibfnamefont {M.}~\bibnamefont
  {{Punk}}},\ }\bibfield  {title} {\enquote {\bibinfo {title} {{Exact solution
  of a two-species quantum dimer model for pseudogap metals}},}\ }\href
  {\doibase 10.1103/PhysRevLett.120.187001} {\bibfield  {journal} {\bibinfo
  {journal} {Phys. Rev. Lett.}\ }\textbf {\bibinfo {volume} {120}},\ \bibinfo
  {pages} {187001} (\bibinfo {year} {2018})},\ \Eprint
  {http://arxiv.org/abs/1712.01854} {arXiv:1712.01854 [cond-mat.str-el]}
  \BibitemShut {NoStop}%
\bibitem [{\citenamefont {{Verheijden}}\ \emph {et~al.}(2019)\citenamefont
  {{Verheijden}}, \citenamefont {{Zhao}},\ and\ \citenamefont
  {{Punk}}}]{Punk19}%
  \BibitemOpen
  \bibfield  {author} {\bibinfo {author} {\bibfnamefont {B.}~\bibnamefont
  {{Verheijden}}}, \bibinfo {author} {\bibfnamefont {Y.}~\bibnamefont
  {{Zhao}}}, \ and\ \bibinfo {author} {\bibfnamefont {M.}~\bibnamefont
  {{Punk}}},\ }\bibfield  {title} {\enquote {\bibinfo {title} {{Solvable
  lattice models for metals with $Z_2$ topological order}},}\ }\href {\doibase
  10.21468/SciPostPhys.7.6.074} {\bibfield  {journal} {\bibinfo  {journal}
  {SciPost Physics}\ }\textbf {\bibinfo {volume} {7}},\ \bibinfo {eid} {074}
  (\bibinfo {year} {2019})},\ \Eprint {http://arxiv.org/abs/1908.00103}
  {arXiv:1908.00103 [cond-mat.str-el]} \BibitemShut {NoStop}%
\bibitem [{\citenamefont {Sachdev}\ \emph {et~al.}(2019)\citenamefont
  {Sachdev}, \citenamefont {Scammell}, \citenamefont {Scheurer},\ and\
  \citenamefont {Tarnopolsky}}]{SS19}%
  \BibitemOpen
  \bibfield  {author} {\bibinfo {author} {\bibfnamefont {S.}~\bibnamefont
  {Sachdev}}, \bibinfo {author} {\bibfnamefont {H.~D.}\ \bibnamefont
  {Scammell}}, \bibinfo {author} {\bibfnamefont {M.~S.}\ \bibnamefont
  {Scheurer}}, \ and\ \bibinfo {author} {\bibfnamefont {G.}~\bibnamefont
  {Tarnopolsky}},\ }\bibfield  {title} {\enquote {\bibinfo {title} {{Gauge
  theory for the cuprates near optimal doping}},}\ }\href {\doibase
  10.1103/PhysRevB.99.054516} {\bibfield  {journal} {\bibinfo  {journal} {Phys.
  Rev. B}\ }\textbf {\bibinfo {volume} {99}},\ \bibinfo {pages} {054516}
  (\bibinfo {year} {2019})},\ \Eprint {http://arxiv.org/abs/1811.04930}
  {arXiv:1811.04930 [cond-mat.str-el]} \BibitemShut {NoStop}%
%%CITATION = ARXIV:1811.04930;%%
\bibitem [{\citenamefont {{Sachdev}}\ \emph {et~al.}(2009)\citenamefont
  {{Sachdev}}, \citenamefont {{Metlitski}}, \citenamefont {{Qi}},\ and\
  \citenamefont {{Xu}}}]{SS09}%
  \BibitemOpen
  \bibfield  {author} {\bibinfo {author} {\bibfnamefont {S.}~\bibnamefont
  {{Sachdev}}}, \bibinfo {author} {\bibfnamefont {M.~A.}\ \bibnamefont
  {{Metlitski}}}, \bibinfo {author} {\bibfnamefont {Y.}~\bibnamefont {{Qi}}}, \
  and\ \bibinfo {author} {\bibfnamefont {C.}~\bibnamefont {{Xu}}},\ }\bibfield
  {title} {\enquote {\bibinfo {title} {{Fluctuating spin density waves in
  metals}},}\ }\href {\doibase 10.1103/PhysRevB.80.155129} {\bibfield
  {journal} {\bibinfo  {journal} {Phys. Rev. B}\ }\textbf {\bibinfo {volume}
  {80}},\ \bibinfo {eid} {155129} (\bibinfo {year} {2009})},\ \Eprint
  {http://arxiv.org/abs/0907.3732} {arXiv:0907.3732 [cond-mat.str-el]}
  \BibitemShut {NoStop}%
\bibitem [{\citenamefont {{Chowdhury}}\ and\ \citenamefont
  {{Sachdev}}(2015)}]{DCSS15b}%
  \BibitemOpen
  \bibfield  {author} {\bibinfo {author} {\bibfnamefont {D.}~\bibnamefont
  {{Chowdhury}}}\ and\ \bibinfo {author} {\bibfnamefont {S.}~\bibnamefont
  {{Sachdev}}},\ }\bibfield  {title} {\enquote {\bibinfo {title} {{Higgs
  criticality in a two-dimensional metal}},}\ }\href {\doibase
  10.1103/PhysRevB.91.115123} {\bibfield  {journal} {\bibinfo  {journal} {Phys.
  Rev. B}\ }\textbf {\bibinfo {volume} {91}},\ \bibinfo {eid} {115123}
  (\bibinfo {year} {2015})},\ \Eprint {http://arxiv.org/abs/1412.1086}
  {arXiv:1412.1086 [cond-mat.str-el]} \BibitemShut {NoStop}%
\bibitem [{\citenamefont {{Scheurer}}\ \emph {et~al.}(2018)\citenamefont
  {{Scheurer}}, \citenamefont {{Chatterjee}}, \citenamefont {{Wu}},
  \citenamefont {{Ferrero}}, \citenamefont {{Georges}},\ and\ \citenamefont
  {{Sachdev}}}]{SCWFGS}%
  \BibitemOpen
  \bibfield  {author} {\bibinfo {author} {\bibfnamefont {M.~S.}\ \bibnamefont
  {{Scheurer}}}, \bibinfo {author} {\bibfnamefont {S.}~\bibnamefont
  {{Chatterjee}}}, \bibinfo {author} {\bibfnamefont {W.}~\bibnamefont {{Wu}}},
  \bibinfo {author} {\bibfnamefont {M.}~\bibnamefont {{Ferrero}}}, \bibinfo
  {author} {\bibfnamefont {A.}~\bibnamefont {{Georges}}}, \ and\ \bibinfo
  {author} {\bibfnamefont {S.}~\bibnamefont {{Sachdev}}},\ }\bibfield  {title}
  {\enquote {\bibinfo {title} {{Topological order in the pseudogap metal}},}\
  }\href {\doibase 10.1073/pnas.1720580115} {\bibfield  {journal} {\bibinfo
  {journal} {Proc. Nat. Acad. Sci.}\ }\textbf {\bibinfo {volume} {115}},\
  \bibinfo {pages} {E3665} (\bibinfo {year} {2018})},\ \Eprint
  {http://arxiv.org/abs/1711.09925} {arXiv:1711.09925 [cond-mat.str-el]}
  \BibitemShut {NoStop}%
\bibitem [{\citenamefont {{Kaul}}\ \emph {et~al.}(2008)\citenamefont {{Kaul}},
  \citenamefont {{Kim}}, \citenamefont {{Sachdev}},\ and\ \citenamefont
  {{Senthil}}}]{Kaul08}%
  \BibitemOpen
  \bibfield  {author} {\bibinfo {author} {\bibfnamefont {R.~K.}\ \bibnamefont
  {{Kaul}}}, \bibinfo {author} {\bibfnamefont {Y.~B.}\ \bibnamefont {{Kim}}},
  \bibinfo {author} {\bibfnamefont {S.}~\bibnamefont {{Sachdev}}}, \ and\
  \bibinfo {author} {\bibfnamefont {T.}~\bibnamefont {{Senthil}}},\ }\bibfield
  {title} {\enquote {\bibinfo {title} {{Algebraic charge liquids}},}\ }\href
  {\doibase 10.1038/nphys790} {\bibfield  {journal} {\bibinfo  {journal}
  {Nature Physics}\ }\textbf {\bibinfo {volume} {4}},\ \bibinfo {pages} {28}
  (\bibinfo {year} {2008})},\ \Eprint {http://arxiv.org/abs/0706.2187}
  {arXiv:0706.2187 [cond-mat.str-el]} \BibitemShut {NoStop}%
\bibitem [{\citenamefont {{Sachdev}}(2019)}]{SS18}%
  \BibitemOpen
  \bibfield  {author} {\bibinfo {author} {\bibfnamefont {S.}~\bibnamefont
  {{Sachdev}}},\ }\bibfield  {title} {\enquote {\bibinfo {title} {{Topological
  order, emergent gauge fields, and Fermi surface reconstruction}},}\ }\href
  {\doibase 10.1088/1361-6633/aae110} {\bibfield  {journal} {\bibinfo
  {journal} {Rep. Prog. Phys.}\ }\textbf {\bibinfo {volume} {82}},\ \bibinfo
  {pages} {014001} (\bibinfo {year} {2019})},\ \Eprint
  {http://arxiv.org/abs/1801.01125} {arXiv:1801.01125 [cond-mat.str-el]}
  \BibitemShut {NoStop}%
\bibitem [{\citenamefont {{Moon}}\ and\ \citenamefont
  {{Sachdev}}(2011)}]{EGMSS11}%
  \BibitemOpen
  \bibfield  {author} {\bibinfo {author} {\bibfnamefont {E.~G.}\ \bibnamefont
  {{Moon}}}\ and\ \bibinfo {author} {\bibfnamefont {S.}~\bibnamefont
  {{Sachdev}}},\ }\bibfield  {title} {\enquote {\bibinfo {title} {{Underdoped
  cuprates as fractionalized Fermi liquids: Transition to
  superconductivity}},}\ }\href {\doibase 10.1103/PhysRevB.83.224508}
  {\bibfield  {journal} {\bibinfo  {journal} {Phys. Rev. B}\ }\textbf {\bibinfo
  {volume} {83}},\ \bibinfo {eid} {224508} (\bibinfo {year} {2011})},\ \Eprint
  {http://arxiv.org/abs/1010.4567} {arXiv:1010.4567 [cond-mat.str-el]}
  \BibitemShut {NoStop}%
\bibitem [{\citenamefont {{Pasquier}}\ and\ \citenamefont
  {{Haldane}}(1998)}]{PH98}%
  \BibitemOpen
  \bibfield  {author} {\bibinfo {author} {\bibfnamefont {V.}~\bibnamefont
  {{Pasquier}}}\ and\ \bibinfo {author} {\bibfnamefont {F.~D.~M.}\ \bibnamefont
  {{Haldane}}},\ }\bibfield  {title} {\enquote {\bibinfo {title} {{A dipole
  interpretation of the $\nu=1/2$ state}},}\ }\href {\doibase
  10.1016/S0550-3213(98)00069-8} {\bibfield  {journal} {\bibinfo  {journal}
  {Nuclear Physics B}\ }\textbf {\bibinfo {volume} {516}},\ \bibinfo {pages}
  {719} (\bibinfo {year} {1998})},\ \Eprint
  {http://arxiv.org/abs/cond-mat/9712169} {arXiv:cond-mat/9712169 [cond-mat]}
  \BibitemShut {NoStop}%
\bibitem [{\citenamefont {Read}(1998)}]{read1998lowest}%
  \BibitemOpen
  \bibfield  {author} {\bibinfo {author} {\bibfnamefont {N.}~\bibnamefont
  {Read}},\ }\bibfield  {title} {\enquote {\bibinfo {title} {{Lowest Landau
  level theory of the quantum Hall effect: The Fermi liquid - like state}},}\
  }\href {\doibase 10.1103/PhysRevB.58.16262} {\bibfield  {journal} {\bibinfo
  {journal} {Phys. Rev. B}\ }\textbf {\bibinfo {volume} {58}},\ \bibinfo
  {pages} {16262} (\bibinfo {year} {1998})},\ \Eprint
  {http://arxiv.org/abs/cond-mat/9804294} {arXiv:cond-mat/9804294 [cond-mat]}
  \BibitemShut {NoStop}%
%%CITATION = COND-MAT/9804294;%%
\bibitem [{\citenamefont {{Xu}}\ and\ \citenamefont {{Sachdev}}(2010)}]{XS10}%
  \BibitemOpen
  \bibfield  {author} {\bibinfo {author} {\bibfnamefont {C.}~\bibnamefont
  {{Xu}}}\ and\ \bibinfo {author} {\bibfnamefont {S.}~\bibnamefont
  {{Sachdev}}},\ }\bibfield  {title} {\enquote {\bibinfo {title} {{Majorana
  Liquids: The Complete Fractionalization of the Electron}},}\ }\href {\doibase
  10.1103/PhysRevLett.105.057201} {\bibfield  {journal} {\bibinfo  {journal}
  {Phys. Rev. Lett.}\ }\textbf {\bibinfo {volume} {105}},\ \bibinfo {eid}
  {057201} (\bibinfo {year} {2010})},\ \Eprint {http://arxiv.org/abs/1004.5431}
  {arXiv:1004.5431 [cond-mat.str-el]} \BibitemShut {NoStop}%
\bibitem [{\citenamefont {{Metlitski}}\ \emph {et~al.}(2015)\citenamefont
  {{Metlitski}}, \citenamefont {{Mross}}, \citenamefont {{Sachdev}},\ and\
  \citenamefont {{Senthil}}}]{MMSS15}%
  \BibitemOpen
  \bibfield  {author} {\bibinfo {author} {\bibfnamefont {M.~A.}\ \bibnamefont
  {{Metlitski}}}, \bibinfo {author} {\bibfnamefont {D.~F.}\ \bibnamefont
  {{Mross}}}, \bibinfo {author} {\bibfnamefont {S.}~\bibnamefont {{Sachdev}}},
  \ and\ \bibinfo {author} {\bibfnamefont {T.}~\bibnamefont {{Senthil}}},\
  }\bibfield  {title} {\enquote {\bibinfo {title} {{Are non-Fermi-liquids
  stable to Cooper pairing?}}}\ }\href {\doibase 10.1103/PhysRevB.91.115111}
  {\bibfield  {journal} {\bibinfo  {journal} {Phys. Rev. B}\ }\textbf {\bibinfo
  {volume} {91}},\ \bibinfo {pages} {115111} (\bibinfo {year} {2015})},\
  \Eprint {http://arxiv.org/abs/1403.3694} {arXiv:1403.3694 [cond-mat.str-el]}
  \BibitemShut {NoStop}%
\bibitem [{\citenamefont {{Chen}}\ \emph {et~al.}(2019)\citenamefont {{Chen}},
  \citenamefont {{Jiang}}, \citenamefont {{Wu}}, \citenamefont {{Lyu}},
  \citenamefont {{Li}}, \citenamefont {{Chittari}}, \citenamefont {{Watanabe}},
  \citenamefont {{Taniguchi}}, \citenamefont {{Shi}}, \citenamefont {{Jung}},
  \citenamefont {{Zhang}},\ and\ \citenamefont {{Wang}}}]{chen2019evidence}%
  \BibitemOpen
  \bibfield  {author} {\bibinfo {author} {\bibfnamefont {G.}~\bibnamefont
  {{Chen}}}, \bibinfo {author} {\bibfnamefont {L.}~\bibnamefont {{Jiang}}},
  \bibinfo {author} {\bibfnamefont {S.}~\bibnamefont {{Wu}}}, \bibinfo {author}
  {\bibfnamefont {B.}~\bibnamefont {{Lyu}}}, \bibinfo {author} {\bibfnamefont
  {H.}~\bibnamefont {{Li}}}, \bibinfo {author} {\bibfnamefont {B.~L.}\
  \bibnamefont {{Chittari}}}, \bibinfo {author} {\bibfnamefont
  {K.}~\bibnamefont {{Watanabe}}}, \bibinfo {author} {\bibfnamefont
  {T.}~\bibnamefont {{Taniguchi}}}, \bibinfo {author} {\bibfnamefont
  {Z.}~\bibnamefont {{Shi}}}, \bibinfo {author} {\bibfnamefont
  {J.}~\bibnamefont {{Jung}}}, \bibinfo {author} {\bibfnamefont
  {Y.}~\bibnamefont {{Zhang}}}, \ and\ \bibinfo {author} {\bibfnamefont
  {F.}~\bibnamefont {{Wang}}},\ }\bibfield  {title} {\enquote {\bibinfo {title}
  {{Evidence of a gate-tunable Mott insulator in a trilayer graphene moir{\'e}
  superlattice}},}\ }\href {\doibase 10.1038/s41567-018-0387-2} {\bibfield
  {journal} {\bibinfo  {journal} {Nature Physics}\ }\textbf {\bibinfo {volume}
  {15}},\ \bibinfo {pages} {237} (\bibinfo {year} {2019})},\ \Eprint
  {http://arxiv.org/abs/1803.01985} {arXiv:1803.01985 [cond-mat.mes-hall]}
  \BibitemShut {NoStop}%
\bibitem [{\citenamefont {{Zhang}}\ and\ \citenamefont
  {{Senthil}}(2019)}]{zhang2019bridging}%
  \BibitemOpen
  \bibfield  {author} {\bibinfo {author} {\bibfnamefont {Y.-H.}\ \bibnamefont
  {{Zhang}}}\ and\ \bibinfo {author} {\bibfnamefont {T.}~\bibnamefont
  {{Senthil}}},\ }\bibfield  {title} {\enquote {\bibinfo {title} {{Bridging
  Hubbard Model Physics and Quantum Hall Physics in Trilayer Graphene/h-BN
  moir{\'e} superlattice}},}\ }\href {\doibase 10.1103/PhysRevB.99.205150}
  {\bibfield  {journal} {\bibinfo  {journal} {Phys. Rev. B}\ }\textbf {\bibinfo
  {volume} {99}},\ \bibinfo {pages} {205150} (\bibinfo {year} {2019})},\
  \Eprint {http://arxiv.org/abs/1809.05110} {arXiv:1809.05110
  [cond-mat.str-el]} \BibitemShut {NoStop}%
\bibitem [{\citenamefont {Scammell}\ \emph {et~al.}(2019)\citenamefont
  {Scammell}, \citenamefont {Patekar}, \citenamefont {Scheurer},\ and\
  \citenamefont {Sachdev}}]{SPSS19}%
  \BibitemOpen
  \bibfield  {author} {\bibinfo {author} {\bibfnamefont {H.~D.}\ \bibnamefont
  {Scammell}}, \bibinfo {author} {\bibfnamefont {K.}~\bibnamefont {Patekar}},
  \bibinfo {author} {\bibfnamefont {M.~S.}\ \bibnamefont {Scheurer}}, \ and\
  \bibinfo {author} {\bibfnamefont {S.}~\bibnamefont {Sachdev}},\ }\bibfield
  {title} {\enquote {\bibinfo {title} {{Phases of SU(2) gauge theory with
  multiple adjoint Higgs fields in 2+1 dimensions}},}\ }\href@noop {} {\
  (\bibinfo {year} {2019})},\ \Eprint {http://arxiv.org/abs/1912.06108}
  {arXiv:1912.06108 [cond-mat.str-el]} \BibitemShut {NoStop}%
%%CITATION = ARXIV:1912.06108;%%
\bibitem [{\citenamefont {Ioffe}\ and\ \citenamefont
  {Larkin}(1989)}]{IoffeLarkin89}%
  \BibitemOpen
  \bibfield  {author} {\bibinfo {author} {\bibfnamefont {L.~B.}\ \bibnamefont
  {Ioffe}}\ and\ \bibinfo {author} {\bibfnamefont {A.~I.}\ \bibnamefont
  {Larkin}},\ }\bibfield  {title} {\enquote {\bibinfo {title} {Gapless fermions
  and gauge fields in dielectrics},}\ }\href {\doibase
  10.1103/PhysRevB.39.8988} {\bibfield  {journal} {\bibinfo  {journal} {Phys.
  Rev. B}\ }\textbf {\bibinfo {volume} {39}},\ \bibinfo {pages} {8988}
  (\bibinfo {year} {1989})}\BibitemShut {NoStop}%
\bibitem [{\citenamefont {{Scheurer}}\ and\ \citenamefont
  {{Sachdev}}(2018)}]{MSSS18}%
  \BibitemOpen
  \bibfield  {author} {\bibinfo {author} {\bibfnamefont {M.~S.}\ \bibnamefont
  {{Scheurer}}}\ and\ \bibinfo {author} {\bibfnamefont {S.}~\bibnamefont
  {{Sachdev}}},\ }\bibfield  {title} {\enquote {\bibinfo {title} {{Orbital
  currents in insulating and doped antiferromagnets}},}\ }\href {\doibase
  10.1103/PhysRevB.98.235126} {\bibfield  {journal} {\bibinfo  {journal}
  {\prb}\ }\textbf {\bibinfo {volume} {98}},\ \bibinfo {eid} {235126} (\bibinfo
  {year} {2018})},\ \Eprint {http://arxiv.org/abs/1808.04826} {arXiv:1808.04826
  [cond-mat.str-el]} \BibitemShut {NoStop}%
\end{thebibliography}%

\end{document}